\documentclass[twocolumn]{aastex63}

\usepackage{graphicx}	
\usepackage{amsmath}	
\usepackage{amssymb}	
\usepackage{makecell}
\usepackage{xcolor}
\usepackage{float}

\newcommand{\beq}{\begin{equation}}
\newcommand{\eeq}{\end{equation}}

\submitjournal{ApJ}

\shorttitle{Exoplanets Orbital Decay}
\shortauthors{Ma \& Fuller}
\graphicspath{{./}{figures/}}

\begin{document}

\title{Orbital Decay of Short-Period Exoplanets via Tidal Resonance Locking}

\correspondingauthor{Linhao Ma}
\email{lma3@caltech.edu}

\author[0000-0001-6117-5750]{Linhao Ma}
\affiliation{TAPIR, Mailcode 350-17, California Institute of Technology, Pasadena, CA 91125, USA}

\author[0000-0002-4544-0750]{Jim Fuller}
\affiliation{TAPIR, Mailcode 350-17, California Institute of Technology, Pasadena, CA 91125, USA}



\begin{abstract}

A large fraction of known exoplanets have short orbital periods where tidal excitation of gravity waves within the host star causes the planets' orbits to decay. We study the effects of tidal resonance locking, in which the planet locks into resonance with a tidally excited stellar gravity mode. Because a star's gravity mode frequencies typically increase as the star evolves, the planet's orbital frequency increases in lockstep, potentially causing much faster orbital decay than predicted by other tidal theories. Due to nonlinear mode damping, resonance locking in Sun-like stars likely only operates for low-mass planets ($M \lesssim 0.1 \, M_{\rm Jup}$), but in stars with convective cores it can likely operate for all planetary masses. The orbital decay timescale with resonance locking is typically comparable to the star's main-sequence lifetime, corresponding to a wide range in effective stellar quality factor ($10^3 \lesssim Q' \lesssim 10^9$), depending on the planet's mass and orbital period. We make predictions for several individual systems and examine the orbital evolution resulting from both resonance locking and nonlinear wave dissipation. Our models demonstrate how  short-period massive planets can be quickly destroyed by nonlinear mode damping, while short-period low-mass planets can survive, even though they undergo substantial inward tidal migration via resonance locking.

\end{abstract}



\section{Introduction} \label{sec:intro}

Historically, exoplanets have been easiest to detect at short orbital periods through transits or radial velocity measurements. Consequently, many known exoplanets orbit at small distances where gravitational forces are strong, allowing the ensuing tidal effects to shape the planetary architectures we observe today. In most cases, the orbits of short-period exoplanets are expected to quickly circularize due to tidal dissipation within the exoplanet, with the spin of the exoplanet aligning and synchronizing with its orbit (though see \citealt{millholland:20} for an exception). Subsequent orbital migration is then driven by tidal dissipation within the star, and it is this case we study here. 

Traditionally, tidal dissipation within the star is parameterized by the effective tidal quality factor $Q'=Q/k_2$, where $Q$ is the inverse of the phase lag between the tidal potential and the tidal bulge \citep{goldreich1966} and $k_2$ is the tidal Love number. In this model, the value of $Q'$ is related to the orbital decay rate by
\beq
\label{eq:Q_define}
Q'\equiv\frac{3M_\mathrm{p}}{M_*}\bigg(\frac{R_*}{a}\bigg)^5t_\mathrm{tide}(\Omega_\mathrm{orb}-\Omega_\mathrm{s})
\eeq
where $M_\mathrm{p}$ and $M_*$ are the masses of the planet and the star, respectively. $R_*$ the radius of the star, $a$ the orbital semi-major axis, $\Omega_\mathrm{orb}$ the angular orbital frequency, $\Omega_\mathrm{s}$ the stellar spin frequency, and $t_\mathrm{tide}$ the tidal migration timescale, defined as
\beq
\label{eq:t_tide_define}
t_\mathrm{tide} = - \frac{a}{\dot{a}_\mathrm{tide}}= \frac{E_{\rm orb}}{\dot{E}_{\rm orb}} \, .
\eeq

Although widely discussed in the literature, $Q'$ is difficult to calculate from first principles and a number of theoretical models have been proposed. Tidal dissipation in exoplanet host stars is believed to result from a combination of a few dissipation mechanisms: (1) damping of the equilibrium tidal distortion of the star via turbulent viscosity in the convective envelope (recent work includes \citealt{mathis:16,gallet:17,duguid:20,vidal:20}),  (2) damping of dynamically excited inertial waves in the convective envelope (e.g., \citealt{papaloizou:10,ogilvie:13,auclair:15,mathis:15,guenel:16}), and (3) thermal and nonlinear dissipation of tidally excited gravity waves in the radiative interior of the star \citep{goodman:98,weinberg:12,ivanov:13,essick:16,fuller2017}. Throughout this paper we will be focusing on gravity wave damping, which is likely to be most effective for planets on circular, short-period orbits aligned with the host star's spin \citep{barker2020}.

Most prior theoretical investigations have overlooked an essential aspect of the tidal migration problem: the coupled evolution of the stellar structure and  the planetary orbit. Even sophisticated models rarely perform full orbital evolution simulations that solve for tidal dissipation at each time step. Instead, they typically invoke a constant tidal quality factor $Q'$, or at best recompute a frequency-averaged $Q'$ at different timesteps. Such averaging is problematic because the effective $Q'$ for gravity waves or inertial waves is a sensitive function of forcing frequency, such that it has sharp minima over narrow frequency ranges surrounding resonances with stellar oscillations. 

In this work, we examine the possibility of tidal migration driven by ``resonance locking" with stellar oscillation modes, in which a planet can become trapped in a resonance with a star's oscillation mode, often allowing for large amounts of tidal dissipation and faster orbital migration. Resonance locking has previously been discussed for binary stars \citep{witte:99,witte:01,fullerkoi54:12,burkart:13,burkart:14,Zanazzi2021}, with direct evidence arising from large-amplitude tidally excited oscillations in eccentric heartbeat stars \citep{hambleton:18,fullerkic81:17,Cheng2020}. Resonance locking within Saturn also appears to drive the orbital expansion of its outer moons Rhea and Titan \citep{fuller2016,lainey:17,Lainey2020} at rates 10-100 times faster than most prior expectations. Resonance locking could have similarly dramatic effects for the inward or outward migration of short-period exoplanets, and we examine this possibility for the first time.

In Section \ref{sec:mechanism} we discuss the tidal dissipation mechanisms, where in Section \ref{sec:resonance_locking} we focus on resonance locking, and in Section \ref{sec:nonlinear} we discuss complications introduced by nonlinear damping effects. We compare our results with observational constraints in Section \ref{sec:comparison_obs}. In Section \ref{sec:discussion}, we discuss the observational implications of resonance locking and other nonlinear tidal theories for exoplanet systems, focusing on individual systems and statistical distributions. We summarize in Section \ref{sec:conclusion}.

\section{Tidal Dissipation Mechanisms} \label{sec:mechanism}

Here we describe the basic idea of resonance locking and the tidal migration time scales it predicts. We also contrast this against tidal migration induced by nonlinear gravity wave dissipation, and we discuss the corresponding tidal $Q'$s and domains of validity of these theories.

\subsection{Resonance Locking}
\label{sec:resonance_locking}

\begin{figure*}
    \includegraphics[width=\textwidth]{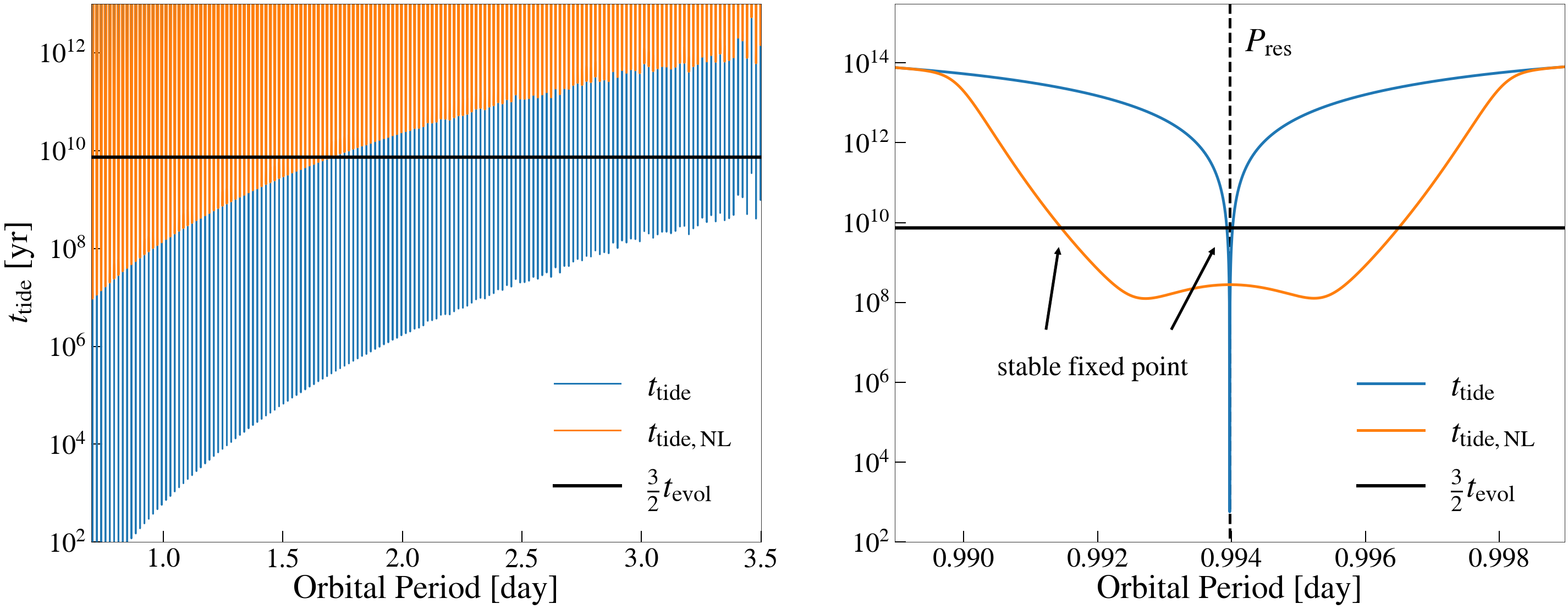}
    \caption{{\bf Left}: tidal migration timescale of a $10\,M_\oplus$ exoplanet as a function of orbital period for a Sun-like star (blue line) due to linear g mode damping, along with a typical mode evolution time scale (black line). The orange line represents a possible modification due to nonlinear damping that saturates g mode resonances. {\bf Right}: zoom-in around the g mode resonance at $P_\mathrm{orb}\approx1\,\mathrm{days}$, showing the stable fixed point where $1.5t_\mathrm{evol}=t_\mathrm{tide}$, corresponding to inward migration via resonance locking. Nonlinear damping makes the resonances shallower, preventing resonance locking at longer periods (see discussion in Section \ref{essick} and Appendix \ref{appendix:nonlin})}
    \label{fig:RL_trapped}
\end{figure*}

Resonances between tidal forcing frequencies and stellar oscillation mode frequencies can greatly enhance tidal dissipation rates. Specifically, the orbital energy loss rate due to a tidally forced mode with angular frequency $\omega_\alpha$ excited by the tidal potential of a circularly orbiting planet with forcing frequency $\omega_\mathrm{f}$ (each measured in a frame corotating with the star) is given by \citep{fuller2017}
\beq
\label{eq:tide_dissip}
\dot{E}_\mathrm{orb,tide} = \frac{m\omega_\alpha \Omega_\mathrm{orb}|\gamma_\alpha| M_*R_*^2|\mathcal{Q}_\alpha|^2\omega_\mathrm{f}^2}{(\omega_\alpha-\omega_\mathrm{f})^2+\gamma_\alpha^2}\bigg(\frac{M_\mathrm{p}}{M_*}\bigg)^2\bigg(\frac{R_*}{a}\bigg)^6\,,
\eeq
where $\gamma_\alpha$ is the mode growth rate and $m$ is the mode's azimuthal index ($m=2$ corresponds to the strongest tidal forcing for aligned orbits). $\mathcal{Q}_\alpha$ is a dimensionless number describing the  spatial coupling between oscillations and the tidal potential defined in \cite{fuller2017}. The denominator is smallest near resonance, when $\omega_\alpha \simeq \omega_\mathrm{f}$, leading to the greatest tidal dissipation and the smallest tidal migration timescale, as shown in the left panel of Figure \ref{fig:RL_trapped}.

\begin{figure*}
    \includegraphics[width=\textwidth]{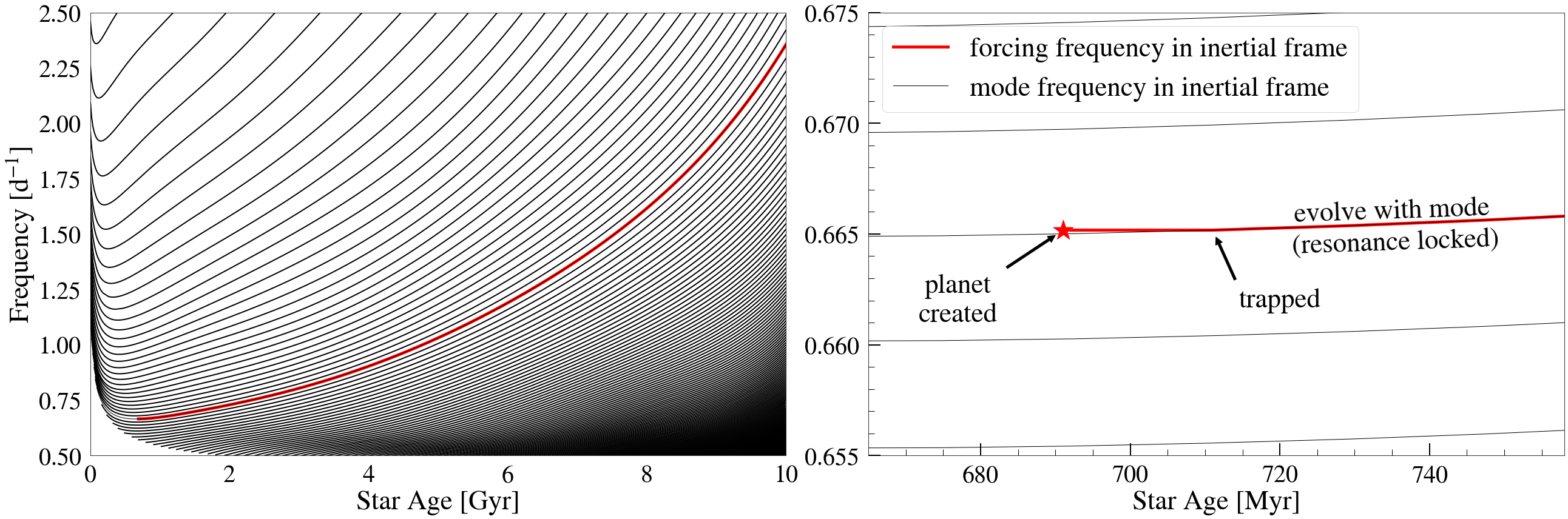}
    \caption{The life of a planet undergoing resonance locking with a $1 \, M_\odot$ star. \textbf{Left}: each line marks the frequency of a stellar g mode in the inertial frame, while the red line is the planet's tidal forcing frequency, which equals twice the orbital frequency in the inertial frame. We only plot one out of three g modes for clarity. A planet born at $\sim \! 700\,\mathrm{Myr}$ in a 3-day orbit will soon get trapped in resonance with one of the modes, causing it to migrate inward via resonance locking.  \textbf{Right}: zoom-in on the moment where resonance locking is first established.}
    \label{fig:RL_explained}
\end{figure*}

The thick radiative zones in main-sequence stars cause internal gravity modes (g modes) to have a dense spectrum in frequency space (see Figure \ref{fig:RL_explained}, left panel). Because the star's internal Brunt-Väisälä frequency typically increases on a stellar evolution timescale, the g mode frequencies increase on a similar time scale, which we define as the mode evolution timescale $t_\alpha\equiv \omega_\alpha/\dot{\omega}_\alpha$. A planet at angular orbital frequency $\Omega_{\rm orb}$ produces tidal forcing at the frequency $\omega_\mathrm{f} = m(\Omega_\mathrm{orb}-\Omega_\mathrm{s})$, where $\Omega_\mathrm{s}$ is the stellar spin frequency. As the stellar oscillation mode frequencies increase, one of them will quickly encounter a resonance with the tidal forcing frequency, i.e., $\omega_\alpha\rightarrow\omega_\mathrm{f}$ (see Figure \ref{fig:RL_explained}, right panel).

As the planet falls into resonance, it can become ``trapped" in resonance (resonantly locked) in the following manner, as shown in Figure \ref{fig:RL_trapped}. If the orbit is perturbed outward such that $\omega_\mathrm{f}$ decreases, it falls deeper into resonance, which increases the tidal dissipation, such that the planet migrates inward and away from exact resonance. If the orbit is perturbed inward such that $\omega_\mathrm{f}$ increases, it moves further from resonance, which decreases the tidal dissipation, allowing the increasing mode frequency to catch up with the planet and sustain the resonant lock. The planet is thus forced to ``ride the mode'' and evolve inwards at the same pace as the mode's resonant location (i.e., a resonance lock), and the planet's orbital frequency increases as the star's oscillation mode frequency increases (see Figure \ref{fig:RL_explained}, right panel). The mode evolution timescale $t_\alpha \equiv \omega_\alpha/\dot{\omega}_\alpha$ hence determines the tidal migration timescale $t_\mathrm{tide}$, which is directly related to $Q'$ by Equation \ref{eq:Q_define}.

\subsubsection{Stellar Models}

To make quantitative predictions, we construct solar-metallicity stellar models with the MESA stellar evolution code \citep{paxton2011,paxton2013,paxton2015,paxton2018,paxton2019}, and we compute their non-adiabatic oscillation modes with the GYRE pulsation code \citep{townsend2013,townsend2018,goldstein2020}. Example inlists are given in the supplementary materials. The models start at zero-age main sequence (ZAMS) with a spin period of 3 days, though we do not include rotational effects within the MESA model. The nontidal angular momentum loss rate is assumed to be similar to Skumanich's law and is calculated via \citep{skumanich1972,Krishnamurthi1997}
\beq
\label{eq:skumanich_law}
\dot{J}_{*,\mathrm{ex}} = K_wI_*\Omega_\mathrm{s}^3\bigg(\frac{M}{M_\odot}\bigg)^{-1/2}\bigg(\frac{R}{R_\odot}\bigg)^{1/2}
\eeq
where $K_w \approx -6\times10^{-12}\,\mathrm{day}$ is a constant fitted by the Sun's spin period and age. 

\begin{figure}
    \includegraphics[width=\columnwidth]{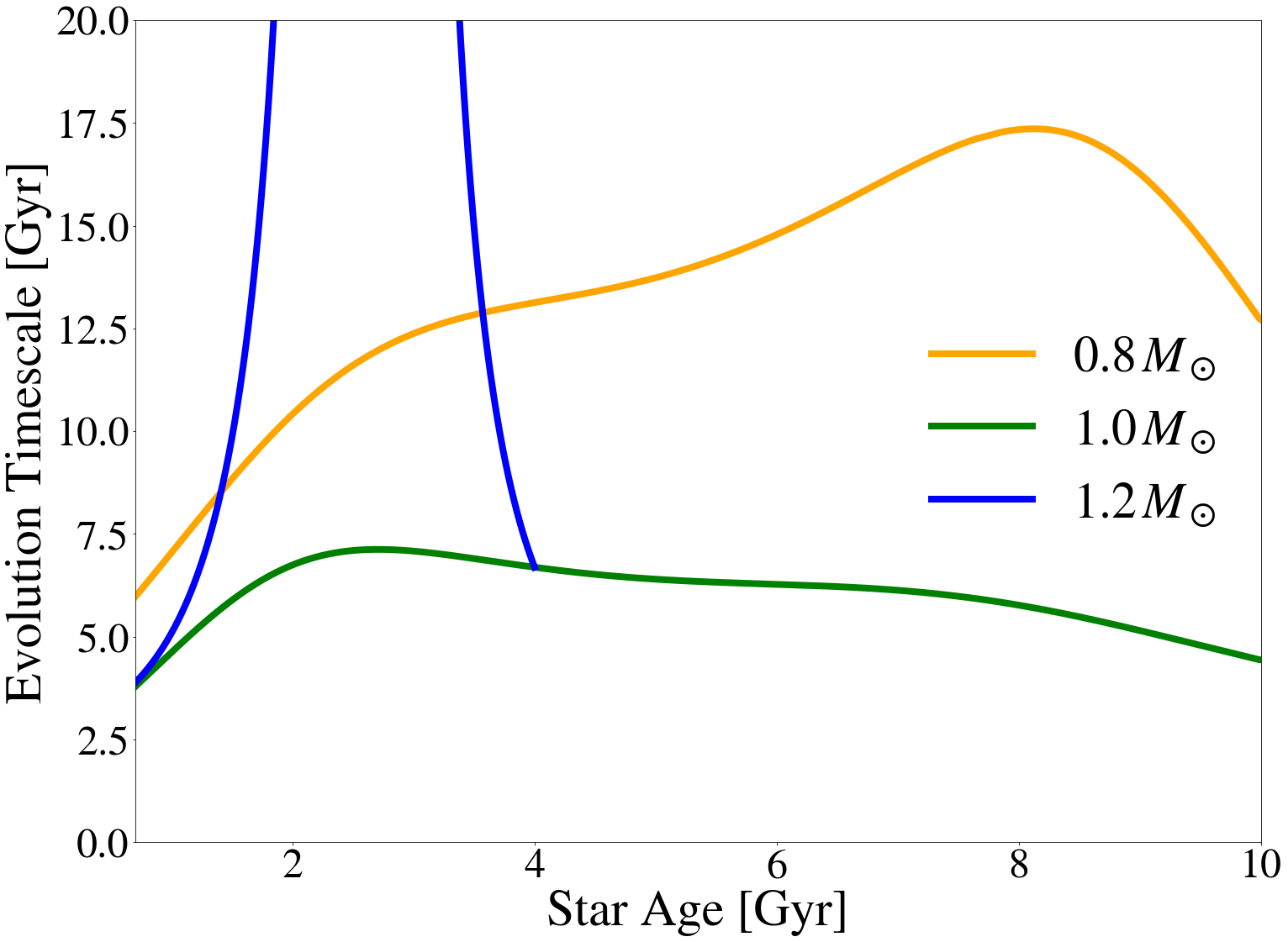}
    \caption{Mode evolution timescales $t_\alpha\equiv\omega_\alpha/\dot{\omega}_\alpha$ of example g modes in $0.8$, $1.0$ and $1.2\,M_\odot$ models. The selected g modes have periods of 1.5 days in the inertial frame at 700 Myr. The mode frequencies typically increase with the star's Brunt-Väisälä frequency as the star evolves. Less massive models have longer mode evolution timescales due to their longer main-sequence lifetime $t_\mathrm{MS}$. The $1.2 \, M_\odot$ model has a negative $t_\alpha$ from about 1.7 to 3.7 Gyr, during which time inward migration via resonance locking cannot occur.}
    \label{fig:T_evol}
\end{figure}

In Figure \ref{fig:T_evol} we plot mode evolution timescales $t_\alpha$ for g modes in stellar models with different masses. The example g modes we plot have periods of 1.5 days in the inertial frame (i.e., they would be resonant with a planet at $P_{\rm orb} = 3\, {\rm days}$) at a stellar age of 700 Myr. We see that $t_\alpha$ is usually comparable to the star's main-sequence lifetime $t_\mathrm{MS}$. The evolution time scale is slightly shorter near the beginning and end of the main sequence when the star's structure changes more rapidly. For the $1.2 \, M_\odot$ model, the g mode frequencies first increase and then decrease with time due to a growing convective core, causing the value of $t_\alpha$ to diverge and then become negative. During that time, inward migration via resonance locking cannot occur because the resonance locations move away from the star rather than toward it.

\subsubsection{Tidal Migration Timescale and Quality Factor}

As discussed above, during resonance locking a star's mode frequency remains nearly equal to the tidal forcing frequency \citep{fuller2016}:
\footnote{
In \cite{fuller2016} $\omega_\alpha = m(\Omega_\mathrm{s}-\Omega_\mathrm{orb})$ during resonance locking. Here we flip the sign for convenience since the stellar spin frequency is usually much smaller than the orbital frequency for short-period exoplanet systems.
}
\beq
\label{eq:freq}
\omega_\alpha \simeq \omega_\mathrm{f} = m(\Omega_\mathrm{orb}-\Omega_\mathrm{s})
\eeq
at all times. Differentiating this equation over time leads to the locking criterion
\beq
\dot{\omega}_\alpha \simeq \dot{\omega}_\mathrm{f} = m(\dot{\Omega}_\mathrm{orb}-\dot{\Omega}_\mathrm{s}) \, .
\eeq
Combining the above equations and defining the spin evolution timescale $t_\mathrm{s} \equiv \Omega_\mathrm{s}/\dot{\Omega}_\mathrm{s}$, we immediately arrive at the (inverse of) the tidal migration timescale
\beq
\label{eq:t_tide_inverse}
t_\mathrm{tide}^{-1} \equiv - \frac{\dot{a}_\mathrm{tide}}{a} = \frac23\frac{\dot{\Omega}_\mathrm{orb}}{\Omega_\mathrm{orb}} = \frac23 \bigg(t_\alpha^{-1} - \frac{\Omega_\mathrm{s}}{\Omega_\mathrm{orb}}(t_\alpha^{-1}-t_\mathrm{s}^{-1})\bigg) \, .
\eeq

However, when tidal migration occurs, the planet adds angular momentum to the stellar spin, which means $t_\mathrm{s}$ and $t_\mathrm{tide}$ are related. This is especially important for systems with massive planets, as shown in \cite{Lainey2020}. Additionally, the system may lose angular momentum due to magnetic braking of the host star. To account for these factors, recall that the total angular momentum $J_\mathrm{tot}$ of the star-planet system is 
\beq
J_\mathrm{tot}=J_*+J_\mathrm{p}=I_*\Omega_\mathrm{s}+M_\mathrm{p}\sqrt{GM_*a} \, ,
\eeq
where $I_*$ is the moment of inertia of the star. Defining the system's change in total angular momentum as $\dot{J}_{*,\rm ex}$, we have
\beq
\label{eq:dot_J}
\dot{J}_{*,\rm ex}=\dot{I}_*\Omega_\mathrm{s}+I_*\dot{\Omega}_\mathrm{s}-\frac12J_\mathrm{p}t_\mathrm{tide}^{-1} \, ,
\eeq
assuming constant stellar/planetary masses as appropriate in most exoplanet systems. If we define evolution timescale for the moment of inertia $t_I=I_*/\dot{I}_*$ and the external stellar spin evolution timescale $t_\mathrm{s,ex}=J_*/\dot{J}_{*,\rm ex}$, this leads us to the relation
\beq
\label{eq:ts}
t_\mathrm{s}^{-1}=t_\mathrm{s,ex}^{-1}-t_I^{-1}+\frac{J_\mathrm{p}}{2J_*}t_\mathrm{tide}^{-1} \, .
\eeq

Substituting Equation \ref{eq:ts} into Equation \ref{eq:t_tide_inverse}, we get the final expression for $t_\mathrm{tide}$
\beq
\begin{split}
\label{eq:t_tide}
t_\mathrm{tide} &= \frac32\frac{\Omega_\mathrm{orb}}{\dot{\Omega}_\mathrm{orb}} \\
&= \frac32 \bigg(1-\frac{I_\mathrm{p}}{3I_*}\bigg)\bigg[\frac{1}{t_\alpha} - \frac{\Omega_\mathrm{s}}{\Omega_\mathrm{orb}}\bigg(\frac{1}{t_\alpha}-\frac{1}{t_\mathrm{s,ex}}+\frac{1}{t_I}\bigg)\bigg]^{-1} \, ,
\end{split}
\eeq
where $I_\mathrm{p} = M_{\rm p} a^2$ is the moment of inertia of the planet's orbit. The pre-factor $1-I_\mathrm{p}/3I_*$ accounts for the angular momentum transport from the planet's orbit to the stellar spin, indicating that the tidal migration timescale becomes very short for massive planets as $I_\mathrm{p}$ approaches $3I_*$, and resonance locking cannot occur if $I_\mathrm{p}>3I_*$. In practice, one can combine Equation \ref{eq:dot_J} and Equation \ref{eq:t_tide} to get a set of coupled differential equations for $\Omega_\mathrm{s}(t)$ and $\Omega_\mathrm{orb}(t)$, and hence solve the full evolution of the spin and orbit numerically. For most short-period exoplanet systems, $\Omega_\mathrm{s}$ is usually negligible compared to $\Omega_\mathrm{orb}$. $I_\mathrm{p}$ is usually negligible compared to $I_*$, except for high-mass or long-period planets.

\begin{figure}
    \includegraphics[width=\columnwidth]{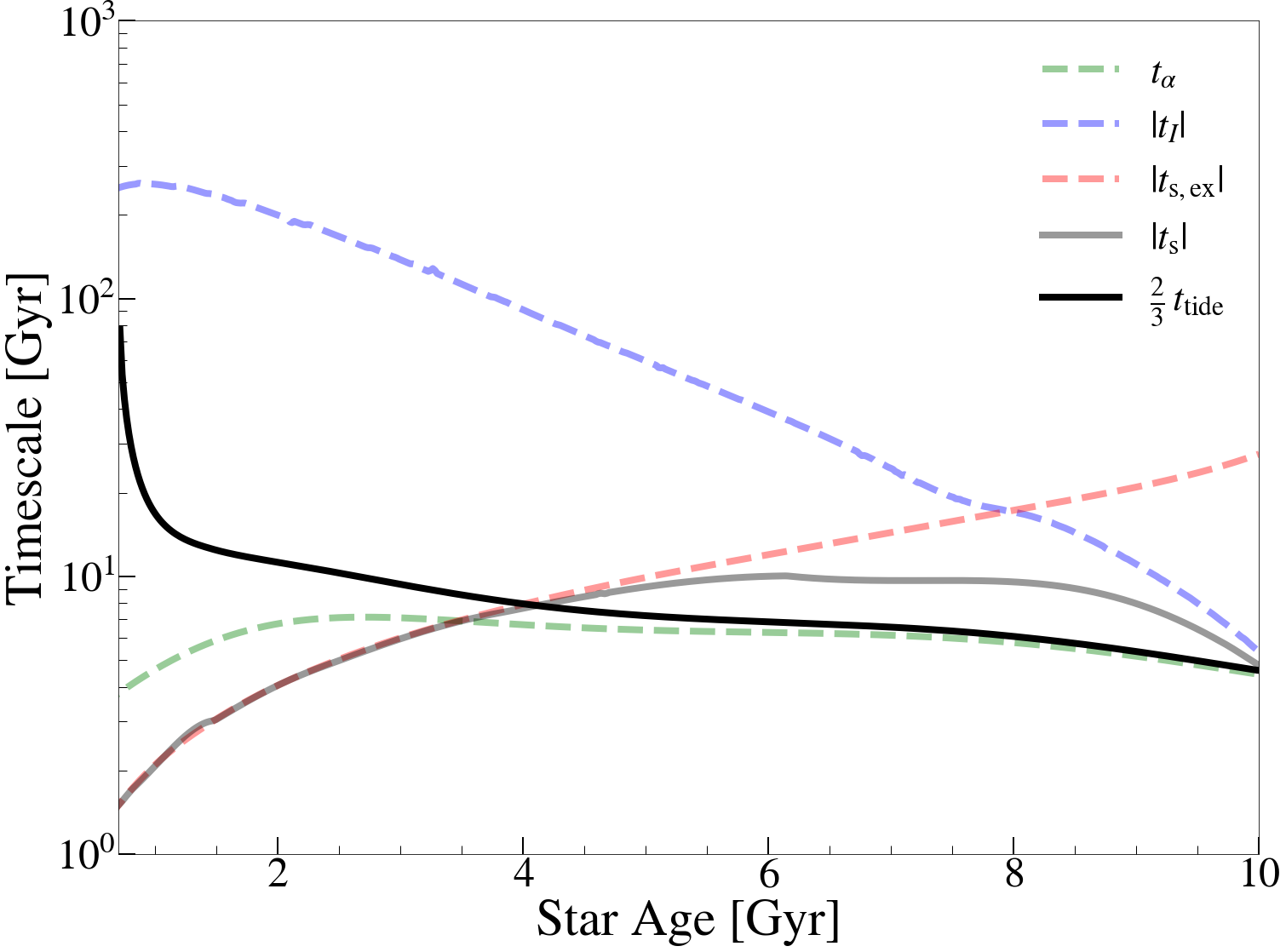}
    \caption{Several related timescales in the expression of $t_\mathrm{tide}$ (Equation \ref{eq:t_tide}) for our $1 \, M_\odot-10\,M_\oplus$ model. Absolute values of negative quantities are plotted. At later ages ($t \gtrsim 3 \, {\rm Gyr}$), stellar rotation is negligible and the shortest timescale is the mode evolution timescale $t_\alpha$, such that $2t_\mathrm{tide}/3 \simeq t_\alpha$ and the planet undergoes orbital decay on a structural evolution time scale. At early ages, rapid stellar spin creates competition between the two terms inside the square bracket of Equation \ref{eq:t_tide}, raising $t_\mathrm{tide}$.}
    \label{fig:T_tide_timescales}
\end{figure}

In Figure \ref{fig:T_tide_timescales}, we plot the relevant evolution timescales for a $1\,M_\odot$ star with a $10\,M_\oplus$ planet. The external spin evolution timescale $t_{s,{\rm ex}}$ is usually comparable to the stellar age, and $t_I$ is always long during the main sequence. At early times when the star is rapidly rotating, the second term in brackets in equation \ref{eq:t_tide} contributes, increasing the value of $t_{\rm tide}$. The star quickly spins down such that $\Omega_s \ll \Omega_{\rm orb}$, at which point $ t_\mathrm{tide} \sim \frac32 t_\alpha$ until the end of the main sequence. Hence, the tidal migration timescale is primarily determined by the evolution timescale of the stellar oscillation mode in resonance with the orbit.

The corresponding effective tidal quality factor during resonance locking is
\beq
\label{eq:RL_Q}
\begin{split}
Q'_\mathrm{RL}&=\frac{9}{2}\frac{M_\mathrm{p}}{M_*}\bigg(\frac{R_*}{a}\bigg)^5\bigg(1-\frac{I_\mathrm{p}}{3I_*}\bigg)\\
\times\,\bigg[\frac{1}{t_\alpha} &- \frac{\Omega_\mathrm{s}}{\Omega_\mathrm{orb}}\bigg(\frac{1}{t_\alpha}-\frac{1}{t_\mathrm{s,ex}}+\frac{1}{t_I}\bigg)\bigg]^{-1}(\Omega_\mathrm{orb}-\Omega_\mathrm{s}) \, .
\end{split}
\eeq
When $\Omega_s \ll \Omega_{\rm orb}$ as appropriate at most stellar ages, this reduces to 
\beq
\label{Qapprox}
\begin{split}
Q'_\mathrm{RL}&\simeq\frac{9}{2}\frac{M_\mathrm{p}}{M_*}\bigg(\frac{R_*}{a}\bigg)^5t_\alpha\Omega_\mathrm{orb}\\
&\approx\frac92\frac{(2\pi)^{13/3}M_\mathrm{p}R_*^5}{G^{5/3}M_*^{8/3}}t_\alpha P_\mathrm{orb}^{-13/3} \, .\\
\end{split}
\eeq
By solving for the evolution of internal oscillation mode frequencies in stellar models, we can quickly compute the corresponding tidal quality factor resulting from resonance locking. Equation \ref{Qapprox} evaluates to
\beq
\label{eq:RL_Q_approx}
\begin{split}
Q'&_\mathrm{RL} \simeq 2\times10^6\times\\
&\bigg(\frac{M_\mathrm{p}}{M_\mathrm{J}}\bigg)\bigg(\frac{M_*}{M_\odot}\bigg)^{\!-8/3}\bigg(\frac{R_*}{R_\odot}\bigg)^{\!5}\bigg(\frac{t_\alpha}{5\,\mathrm{Gyr}}\bigg)\bigg(\frac{P_\mathrm{orb}}{2\,\mathrm{days}}\bigg)^{\!-13/3} \, .
\end{split}
\eeq
That is, $Q'_\mathrm{RL}$ is proportional to the planet mass, and it has a $-13/3$ power-law dependence on the orbital period. This is very different from the prescription of constant $Q'$  that is often assumed in the literature.

\begin{figure*}
    \includegraphics[width=\textwidth]{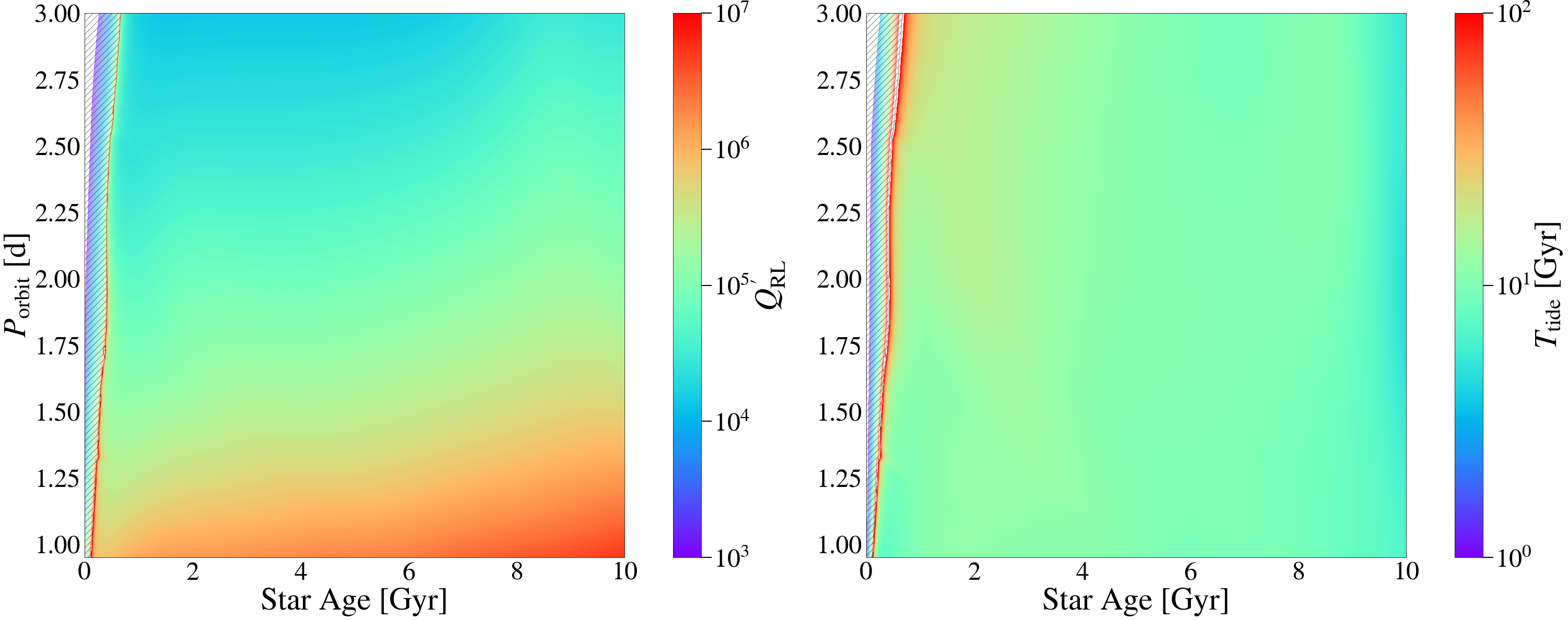}
    \caption{Effective tidal quality factor $Q'_\mathrm{RL}$ (\textbf{left}) and tidal migration timescale $t_\mathrm{tide}$ (\textbf{right}) due to resonance locking between a $1\,M_\odot$ star with a $10\,M_\oplus$ planet. $Q'_\mathrm{RL}$ decreases sharply as the orbital period increases (see Equation \ref{eq:RL_Q}), and increases slowly with time due to the expansion of the star. In contrast, $t_\mathrm{tide}$ remains nearly constant within this parameter space. The primary exception is the red feature at very early ages, which is caused by rapid stellar rotation that creates a divergence in $t_\mathrm{tide}$ (Equation \ref{eq:t_tide}) and $Q'_\mathrm{RL}$. To the left of that feature (hatched regions), $t_{\rm tide}$ is negative and inward migration via resonance locking cannot occur.}
    \label{fig:Q_color}
\end{figure*}

Figure \ref{fig:Q_color} shows the value of $Q'_\mathrm{RL}$ for a $1 \, M_\odot$ star with a $10\,M_\oplus$ planet as a function of orbital period and stellar age. $Q'_\mathrm{RL}$ decreases sharply as the orbital period increases, and it increases somewhat as a function of age primarily because the stellar radius increases slightly, as we would expect from Equation \ref{eq:RL_Q}. In contrast, the value of $t_\mathrm{tide} \simeq \frac32 t_\alpha$ remains nearly constant within this parameter space.

An exception is at early ages, where the stellar spin is larger than a critical frequency such that the second term in the square bracket of Equation \ref{eq:t_tide} is larger than the first, which occurs at approximately
\beq
\label{eq:critical_spin}
\Omega_\mathrm{s} = \Omega_\mathrm{s, crit} \simeq \frac{|t_\mathrm{s,ex}|}{|t_\mathrm{s,ex}|+t_\alpha}\Omega_\mathrm{orb} \, ,
\eeq
where we assumed $t_\mathrm{s}<0$ for main-sequence magnetic braking (spin-down). This would lead to a divergence of $t_\mathrm{tide}$ and $Q'_\mathrm{RL}$. Physically, the divergence signals the boundary where tidal migration due to resonance locking no longer occurs: for a higher spin frequency (or lower orbital frequency) the resonant locations move outward rather than inward. Since $\Omega_\mathrm{s}$ is still less than $\Omega_\mathrm{orb}$ according to Equation \ref{eq:critical_spin}, tidal dissipation would still push the planet inward, and the planet would evolve through the resonances rather than becoming locked in resonance. The colored hatched regions of Figure \ref{fig:Q_color} indicate these regions where resonance locking cannot occur.

At even higher spin frequencies where $\Omega_s > \Omega_{\rm orb}$ (gray regions of Figure \ref{fig:Q_color}), the planet would migrate outward, in the same direction as the resonant locations, such that outward migration via resonance locking could occur. This effect could potentially drive rapid outward migration of short-period planets at very young ages, but we do not study that process in this paper.

\subsection{Nonlinear Wave Dissipation}
\label{sec:nonlinear}

\subsubsection{Validity of linear theory}

The whole theory of resonance locking is based on a linear analysis of dynamical tides \citep{fuller2017}. After the waves get excited at the radiative/convective interface inside a star, they propagate toward the center and are geometrically focused such that their amplitudes increase. We thus expect that resonance locking may not occur if the waves become sufficiently nonlinear near the star's center. Specifically, the dominant nonlinear term in the fluid momentum equation is $\xi\cdot\nabla\xi \sim \xi|d\xi_r/dr|$ for g modes. Hence, the quantity $|d\xi_r/dr|$ naturally serves as a measure of nonlinearity: if $|d\xi_r/dr| \gtrsim 1$, then nonlinear effects become very strong, typically causing wave breaking near the center of the star \citep{barker:11} such that standing g modes no longer exist, and resonance locking cannot occur. In fact, nonlinear g mode damping occurs at smaller g mode amplitudes (see Section \ref{essick}), further limiting the situations in which resonance locking can operate.

\begin{figure}
    \includegraphics[width=\columnwidth]{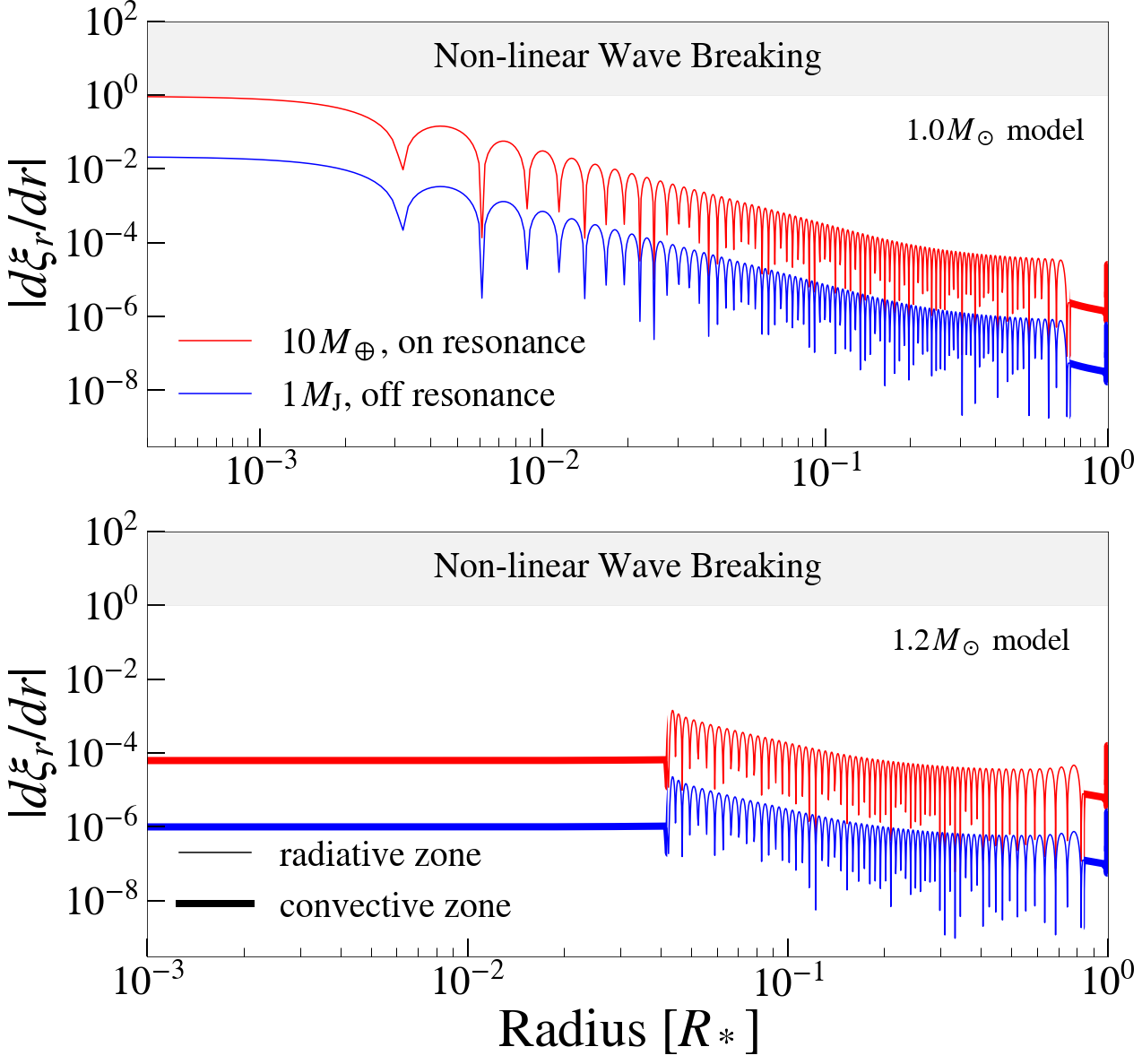}
    \caption{Linearity tests for tidally excited g modes in our $1.0\,M_\odot$ (top panel) and $1.2\,M_\odot$ (bottom panel) models. For each model we compute the value of $|d \xi_r/dr|$ for an off-resonance $1\,M_\mathrm{J}$ hot Jupiter (blue line) and an on-resonance $10\,M_\oplus$ mini-Neptune (red line), both of which are put in a 2-day orbit. We see that g modes in the $1.2\,M_\odot$ model are generally far from wave breaking due to their convective cores (thick lines) which prevent the g modes from propagating near the stellar center. For the $1\,M_\odot$ model with a radiative core, g modes near the stellar center are much more nonlinear but do not reach wave-breaking amplitudes (gray shaded region).}
    \label{fig:linearity}
\end{figure}

When resonance locking does occur, the amplitude of oscillating modes can be calculated as follows: the energy and angular momentum dissipation rates are determined by the tidal migration rate of the planet (Equation \ref{eq:t_tide}). Since energy and angular momentum are conserved, this allows us to compute the corresponding wave amplitude, given a wave damping rate (see \citealt{fuller2017}).\footnote{
Equation \ref{eq:res_lock_amp} follows from \cite{fuller2017} for $N=m=2$ as appropriate for nearly circular orbits.} The result is
\beq
\label{eq:res_lock_amp}
|a_\alpha|_\mathrm{RL} = \frac{1}{2}\bigg|\frac{m\Omega_\mathrm{orb}}{\chi_\alpha\omega_\alpha\gamma_\alpha t_{\alpha,\mathrm{in}}}\bigg|^{1/2}
\eeq
where $\chi_\alpha \equiv 12(M_*+M_\mathrm{p})R^2/(M_\mathrm{p}a^2)-10/(3 \kappa)$ for $l=m=2$ modes and $\kappa\equiv I_*/(M_*R_*^2)$ is the dimensionless moment of inertia of the star. Above, $t_{\alpha,\mathrm{in}} \equiv \sigma_\alpha/\dot{\sigma}_\alpha$ is the mode evolution timescale in the inertial frame, with $\sigma_\alpha = \omega_\alpha+m\Omega_\mathrm{s} \simeq m\Omega_\mathrm{orb}$.

To test the linear approximation for resonantly locked modes in our models, we evaluate the magnitude of $|d \xi_r/dr|$. Figure \ref{fig:linearity} shows $|d\xi_r/dr|$ as a function of radius for a $1.0\,M_\odot$ model at 3.2 Gyr and a $1.2\,M_\odot$ model
\footnote{In the $1.2\,M_\odot$ model, there is a ``jumping core boundary'' issue that prevents us from directly solving the mode evolution timescale. We hence smooth the mode frequency solution by fifth-order polynomials. This is discussed in detail in appendix \ref{appendix:convective_modes}.}
at 1.1 Gyr. For each model we include the mode excited by an off-resonance $1\,M_\mathrm{J}$ hot Jupiter and an on-resonance $10\,M_\oplus$ mini-Neptune (with amplitude calculated via equation \ref{eq:res_lock_amp}), both of which are put in a 2-day orbit. We find that for all models, the g mode nonlinearity indeed increases near the center of the star due to geometrical focusing. However, in the model with a convective core ($1.2\,M_\odot$ model), gravity waves do not propagate into the stellar core, and the g mode always remains linear ($|d \xi/dr| \lesssim 10^{-3}$). For the model with a radiative core ($1.0\,M_\odot$ model), the resonantly locked mode excited by a mini-Neptune comes close to the wave-breaking threshold ($|d \xi/dr| \sim 1$) but does not exceed it. Interestingly, this mode has a larger amplitude than the non-resonant mode excited by a hot Jupiter, demonstrating the great enhancement in amplitude produced by the resonant forcing.

We note that the resonant locking amplitude computed above depends on the damping rate $\gamma_\alpha$ in equation \ref{eq:res_lock_amp}. Weak nonlinear damping may increase the effective value of $\gamma_\alpha$, decreasing the necessary mode amplitude for resonance locking. We revisit this issue in Section \ref{essick}.

\subsubsection{Wave Breaking}

When nonlinear wave breaking occurs near the stellar center, the waves overturn the stratification and are efficiently absorbed \citep{barker:10,barker:11}. The tidally excited gravity waves can then be treated as traveling waves rather than standing g modes, and the corresponding energy dissipation rates have been closely examined in several works 
\citep{zahn:75,goldreich:89,goodman:98,barker2020}.
Specifically, \cite{barker:10} compute the corresponding tidal quality factor for wave breaking,
\beq
\label{eq:barker_Q}
Q'_\mathrm{WB} = 10^5\bigg(\frac{\mathcal{G}_\odot}{\mathcal{G}}\bigg)\bigg(\frac{M}{M_\odot}\bigg)^2\bigg(\frac{R_\odot}{R}\bigg)\bigg(\frac{P_\mathrm{tide}}{0.5\,\mathrm{days}}\bigg)^{8/3}
\eeq
where $P_\mathrm{tide} = 2\pi/\omega_\mathrm{f}$ is the tidal forcing period and $\mathcal{G}$ is a parameter that depends on the stellar structure and is defined in \cite{barker2020}.

Nonlinear wave breaking in Sun-like stars only occurs for planets with $M \gtrsim 3 \, M_{\rm J} (P/1 \, {\rm d})^{-1/6}$, according to \cite{barker:10}. This appears roughly consistent with the calculation of linear mode amplitude in Figure \ref{fig:linearity}. Hence, in our calculations of planetary evolution in Section \ref{sec:discussion}, we only use equation \ref{eq:barker_Q} for the most massive exoplanets.

\subsubsection{Weakly Nonlinear Damping}
\label{essick}

\cite{essick:16} have examined the nonlinear damping of g modes tidally excited by hot Jupiters with periods $P \lesssim 4 \, {\rm days}$ in Sun-like stars. They examined the weakly nonlinear case where the g waves do not break, but they are sufficiently nonlinear to excite daughter and granddaughter modes that dissipate their energy. They found that even for off-resonance hot Jupiters (like the model shown in Figure \ref{fig:linearity}), nonlinear damping is sufficient to wipe out resonances, i.e., the energy dissipation rate is the same for resonant and non-resonant modes. Our on-resonance mini-Neptune in Figure \ref{fig:linearity} excites even larger oscillations than an off-resonance hot Jupiter, meaning that nonlinear damping will dominate over linear damping before the resonant amplitude of equation \ref{eq:res_lock_amp} is reached.

However, \cite{essick:16} also found that nonlinear energy dissipation is much smaller for off-resonance planets with $M_p \lesssim 0.3 \, M_{\rm J}$. For a $10 \, M_\Earth$ planet, this implies that the nonlinear energy dissipation rate is small away from resonance and will greatly increase as the planet moves toward a resonance such that the mode amplitude increases and nonlinear dissipation ramps up. In essence, the total damping rate $\gamma$ in the expression for the mode amplitude is itself a function of the mode amplitude in this situation. If $\gamma$ becomes too large near resonance, the resonance will be ``saturated" (i.e., the blue curve in Figure \ref{fig:RL_trapped} will be moved upward near resonance) such that the resonance locking fixed point does not exist.

To address this possibility, in Appendix \ref{appendix:nonlin} we attempt to extrapolate the results of \cite{essick:16} to low-mass planets below $\simeq 0.3 \, M_{\rm J}$. Using the resulting nonlinear damping rate in place of the linear damping rate in equation \ref{eq:tide_dissip} results in the orange curve in Figure \ref{fig:RL_trapped}, in which the nonlinear damping makes the resonance wells much shallower and wider. The planet may become trapped in resonance if its orbital period is short enough ($P_\mathrm{orb}\lesssim1.7\,\mathrm{days}$ in Figure \ref{fig:RL_trapped}), though we emphasize that more detailed nonlinear coupling calculations are needed for reliable results. This suggests that resonance locking may occur for sufficiently low-mass planets at sufficiently short periods, though the exact mass threshold requires a more accurate calculation of nonlinear damping.

\cite{essick:16} find that the following quality factor provides a good fit to their calculations for sufficiently massive planets around Sun-like stars:
\beq
\label{eq:EW_Q}
Q'_\mathrm{EW} = 2\times10^5 \bigg(\frac{M_\mathrm{p}}{M_\mathrm{J}}\bigg)^{1/2}\bigg(\frac{P_\mathrm{tide}}{0.5\,\mathrm{days}}\bigg)^{2.4}\, .
\eeq
Based on their results, equation \ref{eq:EW_Q} breaks down for planet masses with $M \lesssim 0.3 \, M_{\rm J}$, so we only use this formula for planets in the range $0.3 \, M_{\rm J} \leq M_p \leq 3 \, M_{\rm J}$ in our calculations of orbital evolution in Section \ref{sec:discussion}.

We see that in both Equation \ref{eq:barker_Q} and Equation \ref{eq:EW_Q}, the effective tidal quality factor increases with the orbital period, in stark contrast to the prediction of resonance locking where $Q'$ \textit{decreases} with orbital period. This entails that resonance locking may be more important at longer orbital periods (so long as it can operate), while nonlinear dissipation is likely to dominate at short orbital periods, with important differences in long-term behavior of real systems (see Section \ref{sec:discussion}). 

We conclude that resonance locking will not be prevented by nonlinear effects in stars with convective cores, but nonlinear damping will prevent resonance locking from occurring for hot Jupiters around Sun-like stars. It is unclear whether nonlinear damping will prevent resonance locking of low-mass ($M \lesssim 0.3 \, M_{\rm J}$) planets, and this should be studied in future work.

\section{Comparison with Observations}\label{sec:comparison_obs}

\subsection{Comparison with Penev et al. 2018}\label{sec:relation}

\begin{figure}
    \includegraphics[width=\columnwidth]{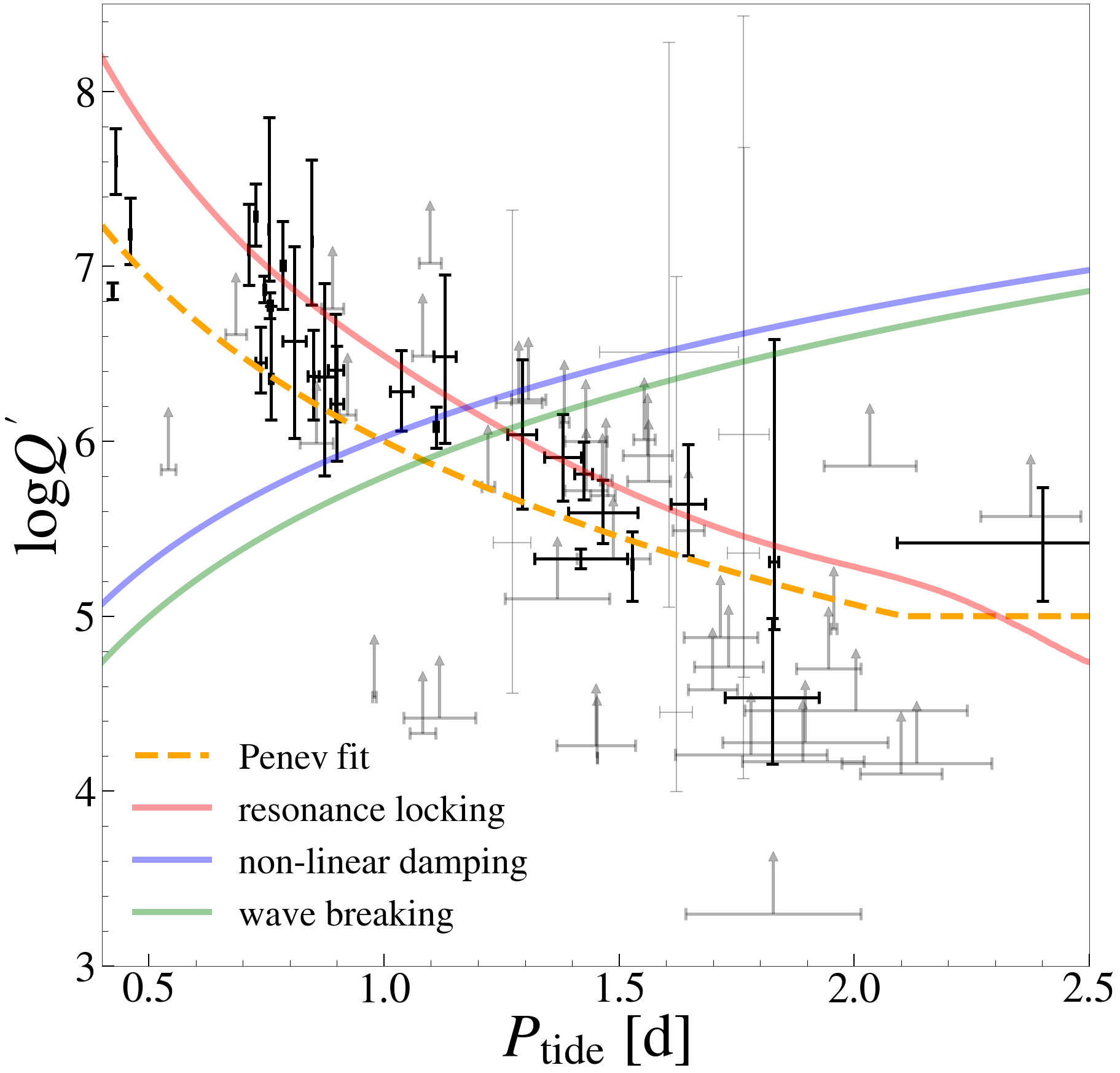}
    \caption{Dependence of $Q^\prime$ on tidal period as predicted by resonance locking (red line), nonlinear damping (blue line) and wave breaking (green line) for models of a Sun-like star with a $1\,M_\mathrm{J}$ planet. We also plot the inferred values of $Q^\prime$ for individual systems from \protect\cite{penev2018}, and their power-law fit (orange dashed line). Black points are cases for which $Q^\prime$ was bounded within two orders of magnitude, while thinner gray symbols are cases with weaker constraints. While the prediction of resonance locking is very similar to the trend from \protect\cite{penev2018}, other explanations for this trend may be possible (see text).}
    \label{fig:Q_T_tide_relation}
\end{figure}

\cite{penev2018} analyzed 188 known hot Jupiter systems to constrain their effect tidal factors based on an improved method from \cite{penev:14}. They managed to constrain two-sided limits on $Q^\prime$ for 35 systems, and to derive lower bounds on $Q^\prime$ for another 40 systems, while the remaining systems in their sample did not lead to meaningful constraints. Of the 75 systems they studied, they found a clear trend toward lower $Q^\prime$ for larger $P_\text{tide}$, where $P_\mathrm{tide}\equiv(P_\mathrm{orb}^{-1}-P_\mathrm{spin}^{-1})^{-1}/m$. The trend is then fitted by the following power-law formula (see Figure \ref{fig:Q_T_tide_relation} or Figure 2 in \citealt{penev2018}):
\beq
\label{eq:relation-fit}
Q' = \max\bigg[10^{6.0} \bigg(\frac{P_\mathrm{tide}}{1 \mathrm{days}}\bigg)^{-3.1},10^5\bigg]
\eeq
When resonance locking occurs and stellar spin is negligible compared to the orbit, Equation \ref{eq:RL_Q_approx} predicts $Q'\approx2\times10^6 \,(P_\mathrm{tide}/1\,\mathrm{day})^{-13/3}$ for fiducial hot Jupiter parameters. This simple analysis immediately leads to a similar power-law trend to the fitted formula \ref{eq:relation-fit}, but with no free parameters. In reality, we expect significant scatter due to the variation of other factors in equation \ref{eq:RL_Q_approx} away from fiducial parameters (e.g., variations in $R_*$ and $M_{\rm p}$ translate to variations in $Q'$).

To compute the exact value of $Q'$ as a function of $P_{\rm tide}$ for resonance locking, we construct a $1\, M_\odot$ stellar model with MESA and compute non-adiabatic oscillation modes. Assuming a stellar rotational evolution based on the modified Skumanich law (Equation \ref{eq:skumanich_law}), we track the run of $Q'$ upon $P_\mathrm{tide}$ for a typical $1\,M_\mathrm{J}$ planet at the stellar age of 5 Gyr, using equation \ref{eq:RL_Q}. The results are shown in Figure \ref{fig:Q_T_tide_relation}. We find that the predicted trend is remarkably similar to the power-law fit by \cite{penev2018}, though slightly offset to higher values of $Q'$. Overall, the resonance locking prediction fits the data very well. We also plot the predicted relations from wave breaking (Equation \ref{eq:barker_Q}) and weakly nonlinear dissipation (Equation \ref{eq:EW_Q}). Those models predict that $Q'$ increases as $P_\mathrm{orb}$ increases, opposite to the trend inferred by \cite{penev2018}.

While these results at first glance appear to provide compelling evidence for the operation of resonance locking in hot Jupiter systems, we caution that other explanations for the trend in $Q'$ from \cite{penev2018} should be examined. From a theoretical perspective, resonance locking likely cannot operate for Jupiter-mass planets around Sun-like stars due to the nonlinear damping discussed in Section \ref{sec:nonlinear}. Hence, we are hesitant to ascribe the trend in $Q'$ from \cite{penev2018} to resonance locking, though resonance locking could provide a nice explanation if nonlinear mode dissipation is much less efficient than found by \cite{essick:16}.

Another possibility worth considering is that the inferred trend in $Q'$ from \cite{penev2018} does not arise from tidal spin-up of the host stars. \cite{penev2018} infer the value of the host stars' $Q'$ by combining age estimates with measurements of spin period. Rapidly rotating host stars (compared to typical field stars) are often inferred to have been tidally spun up by their hot Jupiter companions, entailing that the tidal spin-up time scale is comparable to the main-sequence life time of the host star. This can be seen because much shorter tidal spin-up times would result in the synchronization of the star or the destruction of the hot Jupiter, while much longer ones would not increase the star's rotation significantly. Assuming $t_s = \Omega_{\rm s}/\dot{\Omega}_{\rm s} = 10 \, {\rm Gyr}$, some algebra shows that this requires a scaling $Q' \sim 2 \times 10^6 (P_{\rm tide}/1 \, {\rm day})^{-4} (P_{\rm spin}/10 {\rm days}) (M_{\rm p}/M_{\rm J})^2$, almost identical to the trend shown in Figure \ref{fig:Q_T_tide_relation}. Hence, if one does not have accurate age estimates and assumes that hot Jupiter hosts have similar ages to typical field stars, it could yield a spurious scaling of $Q'$ that is very similar to the trend found by \cite{penev2018}.

Instead, we speculate that moderately rotating host stars of hot Jupiters are (in some cases) simply young stars that are still spinning down, and that they have not been substantially spun up by tides. This may be consistent with the young average ages of hot Jupiter host stars found by \cite{hamer:20a}. In this case, it is very difficult to observationally constrain the host star's $Q'$, except perhaps to place a lower limit. For massive planets where nonlinear effects prevent resonance locking, we expect a trend in $Q'$ similar to that predicted by \cite{essick:16} and \cite{barker2020}. Future work could aim to more accurately constrain the ages of the hot Jupiter host stars in \cite{penev2018} to determine whether their rapid rotation arises from youth or tidal spin-up.

\subsection{Individual Hot Jupiter Systems} \label{sec:system}

Here we summarize our prediction for the effective tidal $Q'$s of 15 real systems based on resonance locking. Most of the systems are chosen from \cite{patra2020}, who argue that the orbital decay of these systems should be the easiest to observe. We also study TRES-3b and WASP-4b which have new observational constraints \citep{bouma2019, mannaday2020}, and we include WASP-128b, a system with a very massive hot Jupiter.

The observational properties of these systems are summarized in Table \ref{tab:systems}. For each system, we construct a number of stellar models to fit their host stars with different initial masses and metallicities within the observational errors and locate the model that matches the other observed properties of the star. We are able to fit the masses, radii, metallicities, ages and effective temperatures within the observational error (see Table \ref{tab:systems} for a summary) for all the host stars except KELT-16b. A typical inlist file is given in the supplementary material.

For each stellar model, we compute $\ell=2$ non-adiabatic oscillation modes with GYRE, with typical inlist files given in the supplementary material. For each system, we assume a negligible spin of the host star and identify the oscillation mode resonant with the tidal forcing, and we calculate the value of $t_\alpha$ for that mode. The effective tidal $Q'$s are then calculated via Equation \ref{eq:RL_Q} using our model properties. We summarize the predictions for resonance locking in Table \ref{tab:system_Qs}, along with predictions for tidal migration rates due to wave breaking \citep{barker2020} and nonlinear dissipation by Equation \ref{eq:EW_Q} \citep{essick:16}. Below are detailed discussions for each system.

\begin{table*}
    \centering
    \begin{tabular}{l|cccclc}
        \hline
        Name & $Q'_\mathrm{obs}$ & $Q'_\mathrm{RL}$ & $Q'_\mathrm{WB}$ & $Q'_\mathrm{EW}$ & Core Status & $t_\mathrm{tide,\,RL}$ (Gyr)\\
        \hline
        HAT-P-23b & $>(3.6\pm1.1)\times10^5$ & $6.0\times10^7$ & ${\bf 3\times10^5}$ &${\bf 4.6\times10^5}$ & radiative? & 5.6\\
        HATS-18b && $2.1\times10^8$ & ${\bf 7\times10^4}$ & ${\bf 1.9\times10^5}$ & radiative & 9.7\\
        KELT-16b & $>(0.5\pm0.1)\times10^5$ & ${\bf 5.1\times10^8}$ & $5\times10^5$ & $3.1\times10^5$ & convective & 9.8\\
        OGLE-TR-56b & $>(4.4\pm1.3)\times10^5$ & ${\bf 1.1\times10^8}$ & $10^6$ & $3.7\times10^5$ & convective & 12.9\\
        TRES-3b & $(5.5\pm4.2)\times10^4$ & $1.2\times10^7$ & ${\bf 4.3\times10^5}$ & ${\bf 5.3\times10^5}$ & radiative & 8.0\\
        WASP-4b & $(1.8\pm0.2)\times10^4$ & $1.1\times10^7$ & ${\bf 2-3\times10^5}$ & ${\bf 4.4\times10^5}$ & radiative & 9.9\\
        WASP-12b & $(1.1\pm0.1)\times10^5$ & $ 1.9\times10^8$ & ${\bf 0.18-3\times10^6}$ & ${\bf 3.0\times10^5}$ & convective? & 7.2\\
        WASP-18b & $>(1.0\pm0.2)\times10^6$ & ${\bf 3.3\times10^8}$ & $2\times10^6$ & $5.8\times10^5$ & convective & 4.7\\
        WASP-19b & $(3.1\pm0.9)\times10^5$ & $2.2\times10^8$ & ${\bf 0.4-0.5\times10^5}$ & ${\bf 1.2\times10^5}$ & radiative & 9.7\\
        WASP-43b & $>(2.5\pm0.2)\times10^5$ & $2.3\times10^8$ & ${\bf 1\times10^5}$ & ${\bf 1.7\times10^5}$ & radiative & 26.6\\
        WASP-72b & $>(1.2\pm0.8)\times10^3$ & $4.5\times10^6$ & $>2\times10^{12}$ & ${\bf 1.7\times10^6}$ & radiative & 0.8\\
        WASP-103b & $>(6.4\pm0.6)\times10^4$ & ${\bf 4.0\times10^8}$ & $2\times10^5$ & $2.1\times10^5$ & convective & 9.7 \\
        WASP-114b && ${\bf 4.5\times10^7}$ & $2\times10^6$ & $7.6\times10^5$ &convective & 7.9\\
        WASP-122b && ${\bf 6.4\times10^6}$ & $2.3\times10^5$ & $8.2\times10^5$ & convective & 2.4\\
        WASP-128b && (no RL) & ${\bf 0.03-1.3\times10^8}$ & ${\bf 8.2\times10^6}$ & radiative & 20.9\\
        \hline
    \end{tabular}
    \caption{Observed/calculated tidal quality factor $Q'$, core status and resonance locking induced $t_\mathrm{tide,\,RL}$ of the systems we study. $Q'_\mathrm{obs}$ shows the observed constraints from \protect\cite{patra2020, bouma2019} and \protect\cite{mannaday2020}. $Q'_\mathrm{RL}$ is the predicted tidal factor from the best-fit resonance locking model. $Q'_\mathrm{WB}$ is the predicted tidal factor from gravity wave breaking \protect\citep{barker2020}, and $Q'_\mathrm{EW}$ is the predicted tidal factor from nonlinear g mode dissipation \protect\citep{essick:16}. We also show the convective/radiative core status of our models. We emphasize that for massive planets like these, resonance locking is only expected to occur in stars with convective cores, and wave breaking/nonlinear dissipation is expected to dominate stars with radiative cores. We bold the $Q'$ values of our inference of the appropriate tidal theory for each system, while the other values are left for reference. The $t_\mathrm{tide,\,RL}$ values show that most system have long tidal migration timescales if they experience resonance locking. {\bf Note}: some authors use a different definition of $Q'$ from ours. We have corrected their results to make them consistent with our definition, so the numbers here may appear different from the original literature.}
    \label{tab:system_Qs}
\end{table*}

\begin{enumerate}

    \item \textbf{HAT-P-23b}: A planet of $2.09\,M_\mathrm{J}$ in a 1.21-day orbit around a G-type dwarf \citep{Bakos2011}. Our best-fit model is a $1.10\,M_\odot$ star with a radiative core. Lack of detected orbital decay requires $Q'>(3.6\pm1.1)\times10^5$ \citep{patra2020}. Our resonance locking calculation predicts $Q'_\text{RL}=6.0\times10^7$, with $t_\mathrm{tide}=5.6\,\mathrm{Gyr}$, but resonance locking is not expected due to nonlinear effects in the radiative core. \cite{barker2020} predicts $Q'_\text{WB}=3\times10^5$ from calculations of wave breaking, but the best-fit model in that work has a convective core. Weak nonlinear mode damping gives $Q'_\text{EW}=4.6\times10^5$. This is a promising system in which to observe tidal decay if the core is indeed radiative.
    
    \item \textbf{HATS-18b}: A planet of $1.98\,M_\mathrm{J}$ in a 0.84-day orbit around a G-type star \citep{penev2016}. Our best-fit model is a $1.03\,M_\odot$ star with a radiative core. No reliable constraint on $Q'$ could be found due to the lack of data \citep{patra2020}. Our resonance locking calculation predicts $Q'_\text{RL}=2.1\times10^8$, with $t_\mathrm{tide}=9.7\,\mathrm{Gyr}$, but resonance locking is not expected due to nonlinear effects since the star has a radiative core and a massive planet. \cite{barker2020} predicts $Q'_\text{WB}\approx7\times10^4$ from calculations of wave breaking, weak nonlinear mode damping predicts $Q'_\text{EW}=1.9\times10^5$. The results make HATS-18b a very promising candidate in which to observe orbital decay.
    
    \item \textbf{KELT-16b}: A planet of $2.75\,M_\mathrm{J}$ in a 0.97-day orbit around an F-type star \citep{oberst2017}. Our best-fit model is a $1.18\,M_\odot$ star with a convective core. Lack of detected orbital decay requires $Q'>(0.5\pm0.1)\times10^5$ \citep{patra2020}. Our resonance locking calculation predicts $Q'_\text{RL}=5.1\times10^8$, with $t_\mathrm{tide}=9.8\,\mathrm{Gyr}$, which is consistent with the observed lower limit, indicating no tidal decay should have been observed.
    
    \item \textbf{OGLE-TR-56b}: A planet of $1.39\,M_\mathrm{J}$ in a 1.21-day orbit around an F-type star \citep{sasselov2003,torres2008}. Our best-fit model is a $1.23\,M_\odot$ star with a convective core. Lack of detected orbital decay requires $Q'>(4.4\pm1.3)\times10^5$ \citep{patra2020}. Our resonance locking calculation predicts $Q'_\text{RL}=1.1\times10^8$, with $t_\mathrm{tide}=12.9\,\mathrm{Gyr}$, which is consistent with the observed lower limit, indicating no tidal decay should be observed.
    
    \item \textbf{TRES-3b}:  A planet of $1.92\,M_\mathrm{J}$ in a 1.306-day orbit around an G-type dwarf \citep{odonovan2007}. Our best-fit model is a $0.89\,M_\odot$ star with a radiative core. Observations indicate $Q'\approx(5.5\pm4.2)\times10^4$ \citep{mannaday2020}, potentially detecting rapid orbital decay. Our resonance locking calculation predicts $Q'_\text{RL}=1.2\times10^7$, with $t_\mathrm{tide}=8.0\,\mathrm{Gyr}$, but resonance locking is not expected due to nonlinear damping in the radiative core. \cite{barker2020} predicts $Q'_\text{WB}=4.3\times10^5$ from calculations of wave breaking, slightly larger than the measured value. Weak nonlinear mode damping predicts $Q'_\text{EW}=5.3\times10^5$. Further observations should attempt to verify the result of \cite{mannaday2020} and will help calibrate models of orbital decay via nonlinear g mode damping in the core.
    
    \item \textbf{WASP-4b}: A planet of $1.186\,M_\mathrm{J}$ in a 1.338-day orbit around a main-sequence star \citep{bouma2019,Southworth2019}. Our best-fit model is a $0.83\,M_\odot$ star with a radiative core. Observations suggest $Q'=(1.8\pm0.2)\times10^4$ \citep{bouma2019}, but \cite{bouma2020} recently discovered a third massive companion that might cause the shift in transit times. Our resonance locking calculation predicts $Q'_\text{RL}=1.1\times10^7$, with $t_\mathrm{tide}=9.9\,\mathrm{Gyr}$, but resonance locking is not expected due nonlinear damping in the radiative core. \cite{barker2020} predicts $Q'_\text{WB}=2-3\times10^5$ from calculations of wave breaking. Weak nonlinear mode damping predicts $Q'_\text{EW}=4.4\times10^5$. Further observations will shed more light on the system and have a good chance of confirming the detection of orbital decay.
    
    \item \textbf{WASP-12b}: A planet of $1.47\,M_\mathrm{J}$ in a 1.09-day orbit around a late F-type main-sequence star or a subgiant \citep{Hebb2009,weinberg2017,collins2017}. Our best-fit model is a $1.44\,M_\odot$ star with a convective core, but sub-giant models without convective cores are also compatible with the data \citep{bailey:19}. Observations indicate orbital decay with a quality factor $Q'=(1.1\pm0.1)\times10^5$ \citep{patra2017,patra2020}. Our resonance locking calculation predicts $Q'_\text{RL}=1.9\times10^8$, three orders of magnitude too high, with $t_\mathrm{tide}=7.2\,\mathrm{Gyr}$. \cite{barker2020} finds $Q'_\text{WB}$ ranging from $1.8\times10^5$ to $3\times10^6$ assuming that gravity waves break near a radiative core, based on different stellar models they choose. Weak nonlinear mode damping predicts $Q'_\text{EW}=3.0\times10^5$. The measured decay rate thus indicates that the star is indeed a sub-giant undergoing orbital decay via nonlinear gravity wave damping.
    
    We speculate that WASP-12b was previously migrating inward slowly via resonance locking when the host star was on the main sequence and had a convective core. When the core became radiative at the end of the main sequence, nonlinear damping became effective, driving the much faster orbital decay we see today. This may help alleviate fine-tuning problems in formation models for WASP-12b, allowing the planet to survive until the end of the main sequence, while also explaining the rapid inward migration at the start of the subgiant phase.
    
    \item \textbf{WASP-18b}: A massive planet of $11.4\,M_\mathrm{J}$ in a 0.94-day orbit around a relatively hot ($T_\mathrm{eff}=6431\,\text{K}$) F-type star \citep{hellier2009,stassun2017}. The mass of the host star is a bit uncertain, with a measurement of $1.46\pm0.29\,M_\odot$ reported. Our best-fit model falls at the low end of the mass measurement, with a mass of $1.17\,M_\odot$ and a convective core. Lack of detected orbital decay requires $Q'>(1.0\pm0.2)\times10^6$ \citep{patra2020}. Our resonance locking calculation predicts $Q'_\text{RL}=3.3\times10^8$, with $t_\mathrm{tide}=4.7\,\mathrm{Gyr}$, which is consistent with the observed lower limit, indicating no tidal decay should have been observed.
    
    \item \textbf{WASP-19b}: A planet of $1.139\,M_\mathrm{J}$ in a 0.79-day orbit around a Sun-like star, making it the hot Jupiter system with the shortest period yet observed \citep{Hebb2010, mancini2013}. Our best-fit model is a $0.91\,M_\odot$ star with a radiative core. Observations indicate $Q'=(3.1\pm0.9)\times10^5$, but the authors encourage caution due to the scanty data \citep{patra2020}. Our resonance locking calculation predicts $Q'_\text{RL}=2.2\times10^8$, with $t_\mathrm{tide}=9.7\,\mathrm{Gyr}$, but resonance locking is not expected due to nonlinear effects in the radiative core. \cite{barker2020} predicts $Q'_\text{WB}\approx4-5\times10^4$ from calculations of wave breaking, smaller than the observational constraint. Weak nonlinear mode damping predicts $Q'_\text{EW}=1.2\times10^5$. Further observations of this system will hence be very useful to constrain tidal theories.
    
    \item \textbf{WASP-43b}: A planet of $2.034\,M_\mathrm{J}$ in a 0.81-day orbit around a K-type dwarf \citep{Hellier2011,gillon2012}. Our best-fit model is a $0.70\,M_\odot$ star with a radiative core. Lack of detected orbital decay requires $Q'>(2.5\pm0.2)\times10^5$ \citep{patra2020}. Our resonance locking calculation predicts $Q'_\text{RL}=2.3\times10^8$, with $t_\mathrm{tide}=26.6\,\mathrm{Gyr}$, but resonance locking is not expected due to nonlinear effects in the radiative core. \cite{barker2020} predicts $Q'_\text{WB}\approx10^5$ from calculations of wave breaking, comparable to the lower limit from observations. Weak nonlinear mode damping predicts $Q'_\text{EW}=1.7\times10^5$. This is another good candidate for orbital decay to be detected in the near future.
    
    \item \textbf{WASP-72b}: A planet of $1.546\,M_\mathrm{J}$ in a 2.22-day orbit around an F-type star \citep{gillon2013}. Our best-fit model is a $1.33\,M_\odot$ subgiant with a radiative core. Lack of detected orbital decay requires $Q'>(1.2\pm0.8)\times10^3$ \citep{patra2020}. Our resonance locking calculation predicts $Q'_\text{RL}=4.5\times10^6$, with $t_\mathrm{tide}=0.8\,\mathrm{Gyr}$, but resonance locking is not expected due to nonlinear damping in the radiative core. \cite{barker2020} predicts a very large $Q'_\text{WB}>10^{12}$ from calculations of wave breaking. However, that work appears to use a stellar model with surface temperature much higher than the observed temperature of $T = 6250 \pm 100 \, {\rm K}$ from \cite{gillon2013}, likely translating to a predicted value of $Q'$ that is far too high. Weak nonlinear mode damping predicts $Q'_\text{EW}=1.7\times10^6$. Future models should re-examine the theoretical predictions.
    
    \item \textbf{WASP-103b}: A planet of $1.51\,M_\mathrm{J}$ in a 0.93-day orbit around a late F-type star \citep{gillon2014,delrez2018}. Our best-fit model is a $1.18\,M_\odot$ star with a convective core. Lack of detected orbital decay requires $Q'>(6.4\pm0.6)\times10^4$ \citep{patra2020}. Our resonance locking calculation predicts $Q'_\text{RL}=4.0\times10^8$, with $t_\mathrm{tide}=9.7\,\mathrm{Gyr}$, which is consistent with the observed lower limit, indicating no tidal decay should have been observed.
    
    \item \textbf{WASP-114b}: A planet of $1.769\,M_\mathrm{J}$ in a 1.55-day orbit around an early G-type star \citep{Barros2016}. Our best-fit model is a $1.24\,M_\odot$ star with a convective core. Being a newly discovered system, no reliable constraint on $Q'$ could be found due to its lack of data \citep{patra2020}. Our resonance locking calculation predicts $Q'_\text{RL}=4.5\times10^7$, with $t_\mathrm{tide}=7.9\,\mathrm{Gyr}$, and the linearity of resonant modes shows resonance locking could occur. Orbital decay is unlikely to be detected for this system unless the star is less massive and contains a radiative core.
    
    \item \textbf{WASP-122b}: A planet of $1.284\,M_\mathrm{J}$ in a 1.71-day orbit around a G-type star \citep{turner2016}. Our best-fit model is a $1.25\,M_\odot$ star with a convective core. Being a newly discovered system, no reliable constraint on $Q'$ could be found due to its lack of data \citep{patra2020}. Our resonance locking calculation predicts $Q'_\text{RL}=6.3\times10^6$, with $t_\mathrm{tide}=2.4\,\mathrm{Gyr}$, and the linearity of resonant modes shows resonance locking could occur. Orbital decay is unlikely to be detected for this system unless the star is less massive and contains a radiative core.
    
    \item \textbf{WASP-128b}: A brown dwarf of $37.19\,M_\mathrm{J}$ in a 2.209-day orbit around an G-type dwarf \citep{Hodzic2018}. Our best-fit model is a $1.13\,M_\odot$ star with a radiative core. No observational constraint on $Q'$ is available at current time. The large companion mass makes the tidal stellar spin-up important, and we do not expect resonance locking in this system because $I_\mathrm{p}>3I_*$ in Equation \ref{eq:t_tide} such that resonance cannot be maintained. \cite{barker2020} predicts $Q'_\text{WB}$ from $3\times10^6$ to $1.3\times10^8$ from calculations of wave breaking, based on the different rotation periods in the stellar models in that work. Weak nonlinear mode damping predicts $Q'_\text{EW}=8.2\times10^6$. Further observations should attempt to measure the stellar spin rate and could potentially detect orbital decay. 
\end{enumerate}

To summarize, for the hot Jupiter systems above, we predict orbital decay timescales of a few gigayears for host stars with convective cores in which resonance locking can operate. Orbital decay via nonlinear mode damping \citep{essick:16} or wave breaking \citep{barker2020} is likely to operate in stars with radiative cores, causing shorter tidal decay time scales that can be more easily observed. Future observations can help confirm our prediction of more rapid orbital decay of hot Jupiters in stars with radiative cores. 

\section{Discussion} \label{sec:discussion}

\begin{figure*}
    \includegraphics[width=\textwidth]{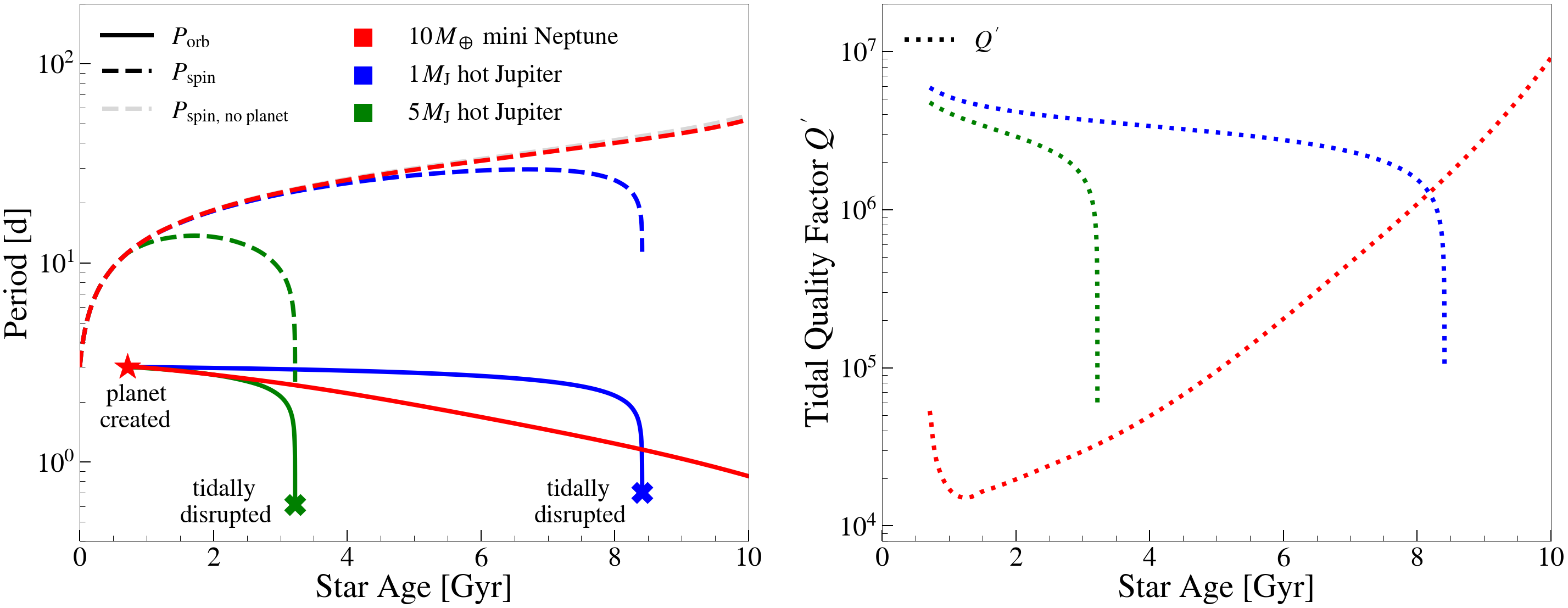}
    \caption{\textbf{Left:} the evolution of planetary orbital period (solid lines) and stellar spin period (dashed lines) for a $1M_\odot$ host star with a $10M_\oplus$ mini-Neptune model (red lines), along with $1M_\mathrm{J}$ (blue lines) and $5M_\mathrm{J}$ (green lines) hot Jupiter models. In the mini-Neptune model, resonance locking is at work, leading to a significant decrease in orbital period during the evolution. Nonlinear mode damping for the $1 \, M_{\rm J}$ planet and wave breaking for the $5 \, M_{\rm J}$ planet lead to tidal disruption on timescales of a few gigayears. \textbf{Right:} the corresponding effective stellar tidal quality factor, $Q'$, over the course of each evolution.}
    \label{fig:full_evolution}
\end{figure*}

\subsection{System Evolution: The Big Picture}
\label{sec:discussion_big_picture}

We have argued that resonance locking, nonlinear g mode dissipation, and gravity wave breaking can all operate in short-period exoplanet systems. In general, one must solve for the angular momentum evolution (equation \ref{eq:dot_J}) with appropriate tidal dissipation physics (i.e. the appropriate value of $Q'$ from equation \ref{eq:Q_define}) to track the full orbital evolution of the system. We expect that resonance locking is the dominant tidal dissipation mechanism for stars massive enough to have convective cores ($M \gtrsim 1.1 \, M_\odot)$, where nonlinear damping is weak and g modes can be resonantly excited. In these stars, equation \ref{eq:t_tide} can be used to estimate the tidal dissipation rate. Heartbeat stars with large-amplitude, tidally excited g modes (e.g., \citealt{fuller2017}) are proof that nonlinear damping does not prevent resonant mode excitation in these stars, even for stellar-mass companions.

For stars with radiative cores, a planet more massive than a few Jupiter masses causes gravity wave breaking near the core \citep{barker:10} such that tidal dissipation is determined by equation \ref{eq:barker_Q}. For Jupiter-mass exoplanets approximately in the range $0.3 \, M_{\rm J} \lesssim M_p \lesssim 3 \, M_{\rm J}$, wave breaking does not occur, but nonlinear mode damping prevents resonant excitation and produces tidal dissipation according to equation \ref{eq:EW_Q} \citep{essick:16}. Resonance locking is likely to be the dominant dissipation mechanism for less massive planets with $M \lesssim 0.3 \, M_{\rm J}$, though future work is needed to quantify this number more accurately (see Section \ref{essick}).

We hence study the orbital evolution of three fiducial exoplanet systems with masses of $10\,M_\oplus, 1\,M_\mathrm{J}$ and $5\,M_\mathrm{J}$, in which resonance locking, nonlinear damping, and wave breaking apply, respectively. We initialize calculations of orbital evolution for planets in 3-day orbits around a $1\,M_\odot$ star at an age of 700 Myr. We integrate the combined equations of orbital decay (equation \ref{eq:t_tide_define}), spin evolution (equation \ref{eq:ts}) and tidal theories (equation \ref{eq:RL_Q}, \ref{eq:barker_Q} or \ref{eq:EW_Q}) to track the full evolution of the systems. We also integrate a system without planets (i.e. a star spinning down purely by magnetic braking) for comparison. The results are shown in Figure \ref{fig:full_evolution}.

For the $10\,M_\oplus$ mini-Neptune, resonance locking causes significant orbital decay during the main sequence, with $t_{\rm tide} \sim t_{\rm MS}$ as typically expected from resonance locking. We note that the effective tidal quality factor driving this planet's inward migration is initially quite small, with $Q' < 10^5$. Consequently, the planet migrates much farther than typical parameterized tidal models with $Q' \sim 10^5-10^6$. Compared to prior work, resonance locking typically predicts substantially more tidal migration for low-mass planets at orbital periods $P \gtrsim 2\, {\rm days}$. In this example, the planet has migrated to ultrashort periods by the end of the main sequence, but it does not plunge into its host star because the value of $Q'$ increases at short periods in order to maintain $t_{\rm tide} \sim t_\alpha$. Due to the relatively low mass of the planet, the stellar spin is hardly influenced by the angular momentum input from the planet's orbit. 

For the $1\,M_\mathrm{J}$ hot Jupiter where nonlinear mode damping dominates the dissipation, the orbit initially decays more slowly than what resonance locking would predict during the first $\sim7\,\mathrm{Gyr}$, because resonance locking would predict a path very similar to that of the $10 \, M_\Earth$ planet above. The slow initial migration is due to the strong period dependence of $Q'$ from equation \ref{eq:EW_Q}, producing an initially large value of $Q'$. However, the orbit decays very rapidly after the planet migrates to a critical period $P \lesssim 2$ days, after which the planet is quickly tidally disrupted. Hence, nonlinear mode coupling predicts rapid orbital decay for hot Jupiters with the shortest periods of $P_{\rm orb} \lesssim 2 \, {\rm days}$, so that such systems are expected to be rare around Sun-like stars with radiative cores. The final plunge also spins up the host star by a factor of $\approx 3$.

A similar evolution occurs for the $5\,M_\mathrm{J}$ planet, which is massive enough to trigger wave breaking. The mass dependence of this mechanism (equation \ref{eq:barker_Q}) entails shorter migration times for more massive planets, so it takes less time ($\sim 3.5\,\mathrm{Gyr}$) for tidal disruption to occur. The host star is highly spun up during the final plunge.

Hence, we conclude that hot Jupiters on short-period orbits around Sun-like stars are likely to be destroyed during the main sequence. Short-period super-Earths and mini-Neptunes are more likely to survive, though their orbits are expected to decay significantly due to resonance locking.

\subsection{Compatibility with Host Star Populations}

Our main predictions appear to be consistent with the recent finding that hot Jupiter host stars are on average slightly younger than field stars \citep{hamer:20a}, implying that a substantial fraction of hot Jupiters are destroyed before their host star evolves off the main sequence. More detailed population modeling will be required to predict an exact number, but we also predict that host stars of short-period (e.g., $P_{\rm orb} \lesssim 3 \, {\rm d}$) hot Jupiters will be younger than host stars of long-period (e.g., $P_{\rm orb} \gtrsim 4 \, {\rm d}$) hot Jupiters. Noting also the well-known trend that higher-mass hot Jupiters have shorter orbital periods on average (e.g., \citealt{owen:18}), we predict that host stars of high-mass hot Jupiters will be younger than host stars of low-mass hot Jupiters, for host stars of nearly the same mass.

In stars with radiative cores, we predict that orbital decay for massive planets ($M \gtrsim 0.3 \, M_{\rm J})$ at short periods ($P \lesssim 2 \, {\rm days}$) proceeds much more rapidly due to nonlinear damping processes. Tidal destruction takes much longer if these processes do not operate (Figure \ref{fig:full_evolution}). Our resonance locking models indicate that wave breaking rarely occurs in stars with convective cores, and the damping produced by nonlinear mode coupling is also likely to be strongly reduced relative to Sun-like stars. Hence, we predict slower orbital decay at short orbital periods and a higher main-sequence survival fraction of hot Jupiters in stars with $M \gtrsim 1.2 \, M_\odot$ than of their lower-mass counterparts.

For mini-Neptunes or less massive planets ($M \lesssim 0.1 \, M_{\rm J}$), resonance locking is probably the dominant tidal dissipation mechanism for stars both with and without convective cores. Resonance locking predicts $t_{\rm tide} \approx 10 \, {\rm Gyr}$ for Sun-like stars regardless of planet mass and orbital period, corresponding to an effective quality factor of 
\beq
Q' \sim 8 \times 10^6 \frac{M_p}{3 M_\oplus} \left( \frac{P_{\rm orb}}{0.5 \, {\rm day}} \right)^{-13/3} \, .
\eeq
This is very close to the inferred constraint from \cite{hamer:20b} for ultrashort-period planets (USPs), though we caution against comparing $Q'$ values because they are very sensitive to the stellar radius. Our predicted tidal migration time scale is comparable to the main-sequence lifetime, consistent with the old ages of USP host stars, but also allowing tidal orbital decay to have significantly shortened the period without destroying the planet.

We can also rule out a naive extrapolation of the nonlinear damping model of \cite{essick:16} to ultrashort-period planets, which predicts a tidal migration time scale of only $t_{\rm tide} \simeq 4 \, {\rm Myr} \left(  M_p/3 M_\oplus \right)^{-0.5} \left( P_{\rm orb}/0.5 \, {\rm d} \right)^{6.7} $, corresponding to a tidal quality factor of $Q' \simeq 6 \times 10^3 \left( M_p/ 3 M_\oplus \right)^{0.5} \left( P_{\rm orb}/0.5 \, {\rm day} \right)^{2.4}$. This is inconsistent with the old ages of host stars from \cite{hamer:20b}, and corresponding inferred lower limits of $Q' \gtrsim 10^7$ for most USPs. Hence, the scaling of the nonlinear migration rate (Equation \ref{eq:EW_Q}) must break down for lower-mass planets as predicted by \cite{essick:16}, likely because the tidally excited gravity modes do not reach sufficient amplitude to transfer energy to daughter modes.

\subsection{System Evolution: Statistical Distributions}

Resonance locking makes unique predictions for the statistical distributions of exoplanet orbits. Consider exoplanets born at a given orbital period $P_\mathrm{i}$ at a constant rate $R=dN(P_\mathrm{i})/dt$, and then migrating inward due to resonance locking/nonlinear dissipation. For a steady-state distribution, the rate of planets migrating through shorter periods is constant, i.e.
\beq
\frac{dN(P)}{dt} = \frac{dN(P_\mathrm{i})}{dt} = R
\eeq
or
\begin{align}
&\frac{dN(P)}{da} = \frac{dN(P)}{dt}\frac{dt}{da} = \frac{R}{\dot{a}}\; \nonumber \\ &\Rightarrow\;\frac{dN(P)}{d\ln P} = \frac23Rt_\mathrm{tide}
\end{align}
For low-mass planets migrating inward via resonance locking, we expect $t_\mathrm{tide}$ is roughly constant, which entails $dN/d\ln P = \mathrm{constant}$, i.e., a uniform distribution over $\log P$. For more massive planets migrating inward via nonlinear wave damping, $t_\mathrm{tide}$ becomes strongly dependent on $P$. For nonlinear mode coupling, this implies
\beq
\frac{dN(P)}{d\ln P} \propto t_\mathrm{tide} \propto P^{6.7} \, ,
\eeq
such that the number of planets should fall very sharply with decreasing orbital period.

However, there are many uncertainties that complicate this simple picture. First, the observed distribution of exoplanets is not necessarily in a steady state. The number of short-period exoplanets may be growing as more numerous exoplanets born at longer periods migrate inward. Second, the birth-period distribution is also likely to be a strong function of period \citep{lee:17}, which complicates interpretation. To first order, we expect resonance locking to shift the birth-period distribution to shorter periods without changing its shape. We therefore expect a flatter distribution of exoplanets at short periods when resonance locking operates, compared to nonlinear dissipation or models with a constant stellar $Q'$ which destroy short-period planets more rapidly.

Recent studies may provide evidence for a distribution of {\em Kepler} planets sculpted by resonance locking migration. For instance, Figures 2 and 3 of \cite{zhu2021} show a relatively uniform occurrence rate of planets with $R_\mathrm{p}\lesssim2R_\oplus$ within the orbital period range $0.6\,\mathrm{days}\lesssim P_\mathrm{orb}\lesssim2\,\mathrm{days}$, which agrees with the basic prediction of resonance locking. In contrast, the occurrence rate of hot Jupiters falls steeply toward short orbital periods, as expected from nonlinear g mode damping for massive planets. A prediction of resonance locking is that the occurrence rate of hot Jupiters should show a flatter trend with orbital period around slightly more massive stars with convective cores. Since resonance locking in individual hot Jupiter systems is generally hard to detect due to the long tidal migration timescale, this may serve as the best prospect to justify whether resonance locking is occurring in these systems. Future population modeling should examine the short-period exoplanetary distribution resulting from resonance locking in more detail.

As resonance locking typically predicts tidal migration timescales $2-3$ orders of magnitudes longer than nonlinear g mode dissipation for hot Jupiters at orbital periods of $\sim 1\, {\rm day}$ (Table \ref{tab:system_Qs}), we might expect short-period hot Jupiters orbiting stars with convective cores (where resonance locking is operating) to be more common than those orbiting Sun-like stars with radiative cores (where nonlinear dissipation dominates). However, several additional factors may complicate this picture: if the hot Jupiters are born at some minimum period (e.g., 3 days), the slow tidal migration induced by resonance locking might prevent them from reaching short orbital periods before the massive star evolves off the main sequence. The observed hot Jupiter population may also suffer from observational biases that preferentially detect systems with certain types of host stars. While disentangling these effects is beyond the scope of this paper, future population analyses  may shed more light on this issue.

\subsection{Early and Late-time evolution}

We have avoided modeling the early evolution ($t \lesssim 500 \, {\rm Myr}$) and post-main-sequence evolution of exoplanet systems in this work. At early times, it is often the case that inward migration via resonance locking cannot occur because resonant locations move outward, as discussed in Section \ref{sec:mechanism}. However, in this case, rapidly rotating young stars with $P_{\rm s} < P_{\rm orb}$ could instead drive outward migration via resonance locking. It is not clear how far planets could be driven outward, but this possibility should be investigated in future work. For example, resonance locking via $m=0$ modes during the pre-main sequence evolution of stellar binaries likely helps to circularize their orbits \citep{Zanazzi2021}.

After the main sequence, the timescales for stellar evolution and those for mode frequency evolution, $t_\alpha$, decrease dramatically, naively resulting in much faster migration via resonance locking. However, post-main-sequence stars contain strongly stratified radiative cores, likely making nonlinear damping in the core even more efficient than in Sun-like stars. Hence, it is not clear whether resonance locking can ever occur in subgiants or stars ascending the red giant branch.

\subsection{Nonlinear damping and the maximum period for resonance locking}

For massive planets, nonlinear damping likely dominates over linear mode damping processes. This causes the resonances to saturate at lower mode amplitudes, decreasing the maximum period $P_{\rm max}$ above which resonance locking cannot operate. The orange line in Figure \ref{fig:RL_trapped} demonstrates this qualitatively, but the crudeness of our approximation of nonlinear damping prevents a quantitative prediction for $P_{\rm max}$. Realistic calculations of nonlinear mode damping rates are needed to reliably predict $P_{\rm max}$. These calculations should be performed for planets of different masses and orbital periods, as well as for stars of different masses and ages. This would allow for better predictions of the statistical distribution of exoplanets as a function of planet mass, orbital period, stellar mass, and stellar age.

\section{Conclusion}
\label{sec:conclusion}

In this work, we study the orbital decay of short-period exoplanets via tidal resonance locking, where planets fall into resonance with stellar oscillation modes and migrate along with the resonant locations (Figure \ref{fig:RL_explained}). When resonance locking between planets and stellar gravity modes (g modes) operates, planetary orbits typically decay on a mode evolution timescale, which is usually similar to the star's main-sequence lifetime. The tidal migration time scale is nearly independent of planet mass and orbital period, such that the effective tidal quality factor $Q'$ decreases toward longer orbital periods and lower-mass planets (equation \ref{eq:RL_Q_approx}).

Resonance locking can be prevented by nonlinear damping that saturates (or eliminates) resonant mode excitation. Both the stellar structure and the planet mass influence the nonlinearity of the tidally excited g modes. For solar-type host stars with radiative cores, nonlinear effects become very important near the center of the star, wiping out resonances. Hot Jupiters of $M \gtrsim 0.3 \,M_\mathrm{J}$ trigger efficient nonlinear dissipation of gravity modes \citep{essick:16}, and more massive planets ($M \gtrsim 3 \, M_{\rm J}$) cause wave breaking \citep{barker2020}. In either case, energy dissipation has a very strong power-law dependence on orbital frequency, with the tidal migration timescale increasing sharply with orbital period. Resonance locking may operate for low-mass planets ($M \lesssim 0.1 \, M_{\rm J}$) around solar-type hosts, and future work should examine this regime. Additionally, resonance locking can likely operate for planets of any mass that orbit massive host stars with convective cores, which prevent gravity waves from reaching the stellar center.

Based on stellar spin measurements, \cite{penev2018} recently inferred a strong period dependence of the tidal quality factor $Q'$ of hot Jupiter host stars (Figure \ref{fig:Q_T_tide_relation}). If resonance locking occurs in hot Jupiter systems, it produces a remarkably similar power-law dependence of $Q'$, which could provide evidence in favor of resonance locking. However, since nonlinear dissipation likely prevents resonance locking from occurring in these systems, other potential explanations should be explored. We have suggested that many moderately rotating hot Jupiter hosts (which were inferred to have been tidally spun up, thereby placing a constraint on $Q'$) are instead simply younger than average \citep{hamer:20a}. In this scenario, their more rapid rotation stems primarily from their youth, and only a lower limit of $Q'$ can be inferred. Future age constraints for those systems may determine which explanation is more likely.

We apply resonance locking to 15 observed hot Jupiter systems and predict that these systems generally have $Q'$s in the range $10^6-10^9$, which is typically $2-3$ orders of magnitude higher than observed lower limits. This means their orbital decay will be hard to measure if resonance locking is operating, as we expect for stars with convective cores. However, nonlinear damping likely operates in host stars possessing radiative cores, leading to much smaller $Q'$s, like that measured for WASP-12b \citep{patra2020}. Further observations of these systems can thus help to improve our understanding of which tidal process operates.

We examine the long-term orbital evolution of exoplanets, combining theories based on resonance locking and nonlinear dissipation/wave breaking (Figure \ref{fig:full_evolution}). We predict that hot Jupiters migrate inwards via nonlinear wave damping and are frequently destroyed during the main sequence for solar-type host stars. This may help to explain the recent finding that hot Jupiter host stars are on average slightly younger than field stars \citep{hamer:20a}. For hot Neptunes and super-Earths, we predict that resonance locking can operate, driving inward migration on a stellar evolution time scale. This can result in a tidal quality factor of $Q' \lesssim10^5$, causing much more orbital decay than prior expectations. However, the corresponding quality factor at short orbital periods can exceed $Q' \gtrsim 10^7$, allowing the planets to survive at ultrashort periods for extended lengths of time, consistent with the observed old ages of ultrashort-period planet hosts \citep{hamer:20b}.

Since nonlinear dissipation occurs for massive planets orbiting stars with radiative cores, we predict a sharp decline in the population of short-period ($P_\mathrm{orb}\lesssim 2 \,\mathrm{days}$) hot Jupiters orbiting solar-type host stars. We predict a more gradual decline for low-mass planets and host stars with convective cores, where resonance locking is at work, producing a much smoother distribution with orbital period. Future observations will help test this prediction, provided that effects of tidal migration can be distinguished from the birth-period distribution (e.g., \citealt{lee:17}).

\acknowledgments
We thank Hang Yu, Rich Townsend, and Josh Winn for very helpful discussion and feedback. This work is partially supported by NASA through grant 20-XRP20 2-0147. JF is thankful for support through an Innovator Grant from The Rose Hills Foundation, and the Sloan Foundation through grant FG-2018-10515.


\software{MESA \citep{paxton2011,paxton2013,paxton2015,paxton2018,paxton2019}, GYRE \citep{townsend2013,townsend2018,goldstein2020}}

\clearpage



\appendix

\section{Table of observed and modelled system properties}

See Table \ref{tab:systems} for details.
\begin{table*}
    \centering
    \setlength{\tabcolsep}{10pt}
    \begin{tabular}{l|ccccccc}
        \hline
        Name & $M_*/M_\odot$ & $R_*/R_\odot$ & $T_\mathrm{eff}/\mathrm{K}$ & age/Gyr & [Fe/H] & $M_\mathrm{p}/M_\mathrm{J}$ & $P/\mathrm{day}$\\
        \hline
        HAT-P-23b$^1$ & 1.13(4) & 1.20(7) & 5905(80) & 4.0(1.0) & 0.15(4) & 2.09(11) & 1.21 \\
        (model) & 1.102 & 1.25 & 5985 & 4.74 & 0.176 && \\
        \Xhline{.5\arrayrulewidth}
        HATS-18b$^2$ & 1.04(5) & 1.02(6) & 5600(120) & 4.2(2.2) & 0.28(8) & 1.980(77)  & 0.84 \\
        (model) & 1.032 & 1.02 & 5627 & 5.6 & 0.337 & & \\
        \Xhline{.5\arrayrulewidth}
        KELT-16b$^3$ & 1.21(5) & 1.36(6) & 6236(54) & 3.1(3) & -0.002(90) & 2.75(16)  & 0.97 \\
        (model) & 1.183 & 1.42 & 6197 & 2.7 & 0.074 & &  \\
        \Xhline{.5\arrayrulewidth}
        OGLE-TR-56b$^4$ & 1.23(8) & 1.36(9) & 6050(100) & 3.1(1.2) & 0.22(10) & 1.39(18) & 1.21 \\
        (model) & 1.227 & 1.38 & 6032 & 3.0 & 0.283 & & \\
        \Xhline{.5\arrayrulewidth}
        TRES-3b$^5$ & 0.90(15) & 0.80(5) & 5720(150) &&& 1.92(23) & 1.306 \\
        (model) & 0.889 & 0.82 & 5570 & 1.52 & 0.0 & & \\
        \Xhline{.5\arrayrulewidth}
        WASP-4b$^6$ & 0.86(18) & 0.89(7) & 5400(180) & 7.0(2.0) & -0.07(38) & 1.19(20)  & 1.338 \\
        (model) & 0.825 & 0.83 & 5573 & 7.23 & -0.078 & & \\
        \Xhline{.5\arrayrulewidth}
        WASP-12b$^7$ & 1.43(1) & 1.66(5) & 6360(140) & 2.0(1.0) & 0.33(17) & 1.470(76)  & 1.09 \\
        (model) & 1.435 & 1.70 & 6238 & 1.8 & 0.337 & & \\
        \Xhline{.5\arrayrulewidth}
        WASP-18b$^8$ & 1.46(29) & 1.29(5) & 6431(48) & 1.0(5) & 0.00(9) & 11.4(1.4)  & 0.94 \\
        (model) & 1.172 & 1.25 & 6408 & 1.0 & -0.027 & & \\
        \Xhline{.5\arrayrulewidth}
        WASP-19b$^9$ & 0.94(4) & 1.02(1) & 5460(90) & 10.2(3.8) & 0.14(11) & 1.139(36)  & 0.79 \\
        (model) & 0.906 & 1.02 & 5510 & 13.2 & 0.237 & &  \\
        \Xhline{.5\arrayrulewidth}
        WASP-43b$^{10}$ & 0.72(3) & 0.67(1) & 4520(120) && -0.01(12) & 2.034(52)  & 0.81 \\
        (model) & 0.696 & 0.67 & 4560 & 6.75 & 0.085 & &  \\
        \Xhline{.5\arrayrulewidth}
        WASP-72b$^{11}$ & 1.39(6) & 1.98(24) & 6250(100) & 3.2(6) & -0.06(9) & 1.546(59) & 2.22 \\
        (model) & 1.331 & 2.19 & 6347 & 2.78 & 0.028 & & \\
        \Xhline{.5\arrayrulewidth}
        WASP-103b$^{12}$ & 1.21(11) & 1.42(4) & 6110(160) & 4.0(1.0) & 0.06(13) & 1.51(11)  & 0.93 \\
        (model) & 1.179 & 1.46 & 6163 & 3.0 & 0.063 & & \\
        \Xhline{.5\arrayrulewidth}
        WASP-114b$^{13}$ & 1.29(5) & 1.43(60) & 5940(140) & 4.0(2.0) & 0.14(7) & 1.769(64)  & 1.55 \\
        (model) & 1.244 & 1.54 & 6079 & 3.08 & 0.178 &&\\
        \Xhline{.5\arrayrulewidth}
        WASP-122b$^{14}$ & 1.24(4) & 1.52(3) & 5720(130) & 5.11(80) & 0.32(9) & 1.284(32)  & 1.71 \\
        (model) & 1.252 & 1.53 & 5713 & 5.03 & 0.406 & & \\
        \Xhline{.5\arrayrulewidth}
        WASP-128b$^{15}$ & 1.16(8) & 1.15(4) & 5950(100) & 2.2(1.8) & 0.01(24) & 37.19(1.70) & 2.209 \\
        (model) & 1.127 & 1.15 & 6019 & 2.52 & 0.207 & & \\
        \hline
    \end{tabular}
    \caption{Properties of the systems we study. For each system, the first line shows the values inferred from observational literature, with numbers in brackets corresponding to 95\% confidence intervals. The second line shows the parameters of our best-fit MESA models. Planetary parameters ($M_\mathrm{p}$ and $P$) are taken directly from the literature. References for observations: 1. \protect\citealt{Bakos2011}, 2. \protect\citealt{penev2016}, 3. \protect\citealt{oberst2017} , 4. \protect\citealt{torres2008}, 5. \protect\citealt{odonovan2007}, 6. \protect\citealt{bouma2019}, 7. \protect\citealt{collins2017}, 8. \protect\citealt{hellier2009}, 9. \protect\citealt{mancini2013}, 10. \protect\citealt{gillon2012}, 11. \protect\citealt{gillon2013}, 12. \protect\citealt{delrez2018}, 13. \protect\citealt{Barros2016}, 14. \protect\citealt{turner2016}, 15. \protect\citealt{Hodzic2018}.}
    \label{tab:systems}
\end{table*}

\section{Solving Modes for MESA Models with Convective Cores}
\label{appendix:convective_modes}

Throughout the paper, we have constructed MESA models to track the evolution of the stellar structure. We then use GYRE to solve the stellar oscillation modes for individual profiles generated by MESA and study the evolution of the modes by tracking the same mode across different profiles at different stellar ages. While this process is straightforward for models of Sun-like stars, it frequently fails for models of massive stars with convective cores.

\begin{figure}
    \includegraphics[width=\columnwidth]{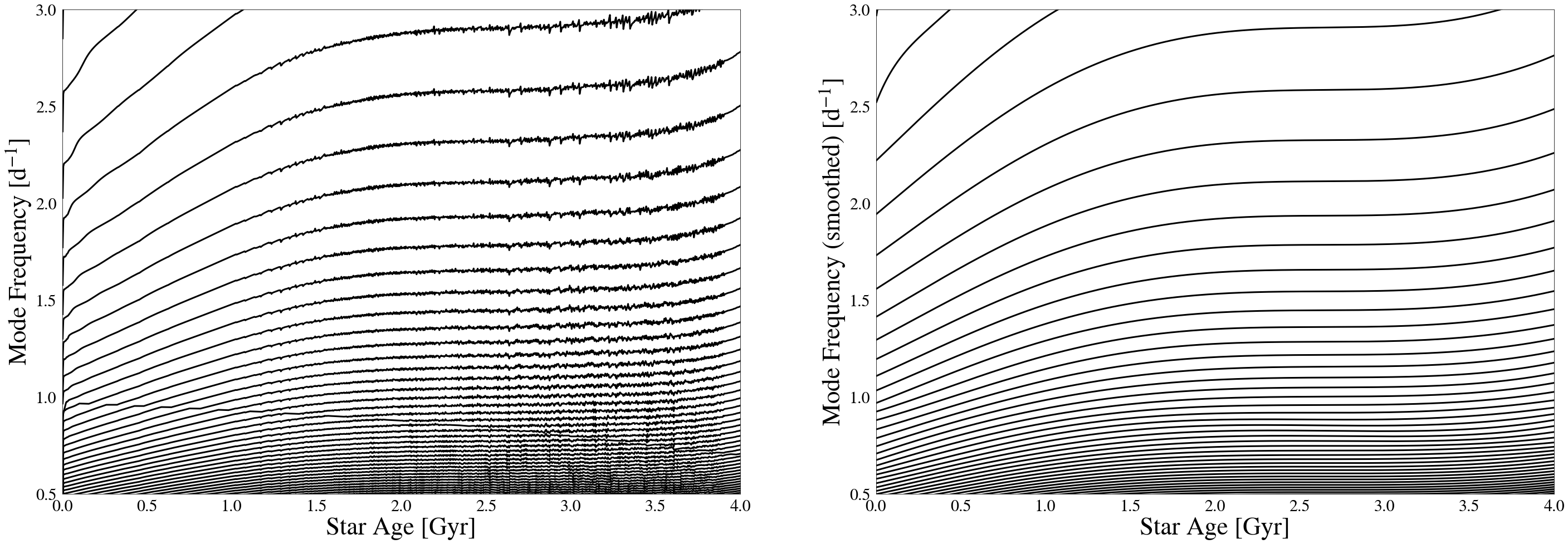}
    \caption{Mode frequencies of a $1.2\,M_\odot$ MESA model. For models with convective cores, it is generally difficult for MESA to accurately determine the boundary of the convective core. This can cause unphysical jumping in the computed mode frequencies (left panel). We smooth the mode frequencies in time by fitting fifth-order polynomials (right panel), giving a more accurate estimate of mode evolution timescales.}
    \label{fig:smooth_compare}
\end{figure}

In Figure \ref{fig:smooth_compare} we describe what we refer to the ``jumping core boundary'' issue for models with convective cores. It is generally difficult for MESA to accurately determine the position of convective core boundaries in the presence of composition gradients. As a result, the mode frequencies solved by GYRE exhibit unphysical jumping, due to the discontinuous jumps in the core boundary. We find that turning on predictive mixing and element diffusion in MESA, as well as choosing smaller time steps and mesh spacing, helps to decrease the unphysical jumping (as shown in Figure \ref{fig:smooth_compare}, left panel). However, this is still not satisfactory when solving for mode evolution timescales, which is related to the derivatives of the frequency, so that even small jumps in the frequencies result in large errors.

Therefore, we choose an alternative approach to determine the mode evolution timescale. Instead of solving the derivatives directly, we fit the frequency solutions with fifth-order polynomials (as shown in Figure \ref{fig:smooth_compare}, right panel). This enables us to compute smoothly varying mode evolution time scales $t_\alpha$, as shown in Figure \ref{fig:T_evol}.

\section{Estimate of nonlinear damping rate}
\label{appendix:nonlin}

Nonlinear mode damping can be modeled as an additional amplitude-dependent damping term $\gamma_{\rm NL}$. The increased damping will cause the Lorentzian dips in $t_{\rm tide}$ in Figure \ref{fig:RL_trapped} to become broader and shallower, altering where resonance locking can operate.

To estimate the nonlinear damping rate, we first realize that the maximum damping rate achievable is the rate at which waves propagate from the convective envelope (where they are excited) to the center of the star (where they are dissipated). This damping rate is the inverse group travel time, $\gamma_{\rm NL,max} \sim -1/\tau_2$, where
\begin{align}
\tau_2 &= \int_{r_0}^{r_c} \frac{dr}{v_g} \nonumber \\
&=\frac{\sqrt{6}}{ \omega^2} \int_{r_0}^{r_c} \frac{dr}{r} N \,  
\end{align}
where we have used the g mode dispersion relation $\omega^2 = N^2\ell(\ell+1)/k^2r^2$ where $\ell=2$ is the mode's spherical harmonic index for tidally excited gravity waves. $r_0 \simeq 0$ is the inner turning point, $r_c$ is the outer turning point at the base of the convective envelope. We note that $\tau_2 = \int^{r_c}_{r_0}kdr/\omega = n\pi$, hence $\tau_2$ scales with the frequency spacing $\Delta\omega_g$ as $\tau_2=\pi/\Delta\omega_g$.

The maximum damping rate $\gamma_{\rm NL,max} \sim -1/\tau_2$ will be achieved for modes with large enough amplitude, which dissipate efficiently after one wave crossing time. Nonlinear wave breaking can be approximated by this damping rate, but sufficiently strong three-mode coupling could produce the same effective damping rate. Modes at smaller amplitudes $a$ will be damped at smaller rates. In Sun-like stars, nonlinear g mode damping is caused by a nonlinear instability in which daughter modes are driven to larger amplitude by the tidally excited parent mode \citep{kumar:96,weinberg:12}. The instability only occurs above a threshold amplitude $a_{\rm NL}$, hence we expect very little nonlinear damping below this threshold. The nonlinear damping rate should fall sharply for $|a| \lesssim |a_{\rm NL}|$, hence we model the nonlinear damping via an ad hoc relation
\begin{equation}
\gamma_{\rm NL} \sim - \frac{1}{\tau_2} \exp \bigg(- \bigg(\frac{a_{\rm NL}}{a}\bigg)^2 \bigg) \, .
\end{equation}
Defining the dimensionless parameter $\bar{\gamma} = -\tau_2\gamma_\mathrm{NL}>0$. we have $(a_\mathrm{NL}/a)^2 = -\ln(\bar{\gamma})$.

To estimate the value of $a_{\rm NL}$, we examine the results of \cite{essick:16} for Sun-like stars. They find that the orbital decay rate for off-resonance modes is weakly dependent on planet mass (and hence mode amplitude) for planets with mass $M_\mathrm{p} \gtrsim 0.3 \, M_{\rm J}$, while the energy dissipation rate is strongly amplitude-dependent for $M_\mathrm{p} \lesssim 0.3 \, M_{\rm J}$. There appears to be a very weak dependence of this cutoff on orbital period, as we might expect since the g mode nonlinearity scales as $|k_r \xi_r| \propto P^{1/6}$. Therefore, their results suggest that $a \simeq a_{\rm NL}$ for planets with $M \simeq 0.3 \, M_{\rm J}$ and resonant detuning $|(\omega_\alpha - \omega_{\rm f})| = \Delta \omega \simeq \Delta \omega_g/2$. Since the tidally excited mode amplitude scales as $a \propto M_p$ and $a \propto (\Delta \omega^2+\gamma^2)^{-1/2}$, we expect at low amplitudes that 
\begin{equation}
\ln(\bar{\gamma}) \sim - \bigg(\frac{0.3 \, M_{\rm J}}{M_\mathrm{p}}\bigg)^2 \bigg( \frac{\Delta \omega^2+(\gamma_\mathrm{rad}+\gamma_\mathrm{NL})^2}{(\Delta \omega_g/2)^2+(\gamma_\mathrm{rad}+1/\tau_2)^2}\bigg)\, .
\end{equation}
Near resonance, the nonlinear damping is expected to be strong such that $\gamma_\mathrm{rad}$ can be neglected. With $\Delta\omega_g\tau_2 = \pi$ we have
\beq
\bar{\gamma}^2 + (1+(\pi/2)^2)\bigg(\frac{M_\mathrm{p}}{0.3 \, M_{\rm J}}\bigg)^2\ln(\bar{\gamma}) + \tau_2^2\Delta \omega^2 = 0\, .
\eeq
When $M_\mathrm{p}\ll\mathrm{0.3\,M_\mathrm{J}}$, we expect $\bar{\gamma}^2=\exp(-2(a_\mathrm{NL}/a)^2)$ to be exponentially smaller than $(M_\mathrm{p}/0.3\,M_\mathrm{J})^2\ln(\bar{\gamma})\sim (M_\mathrm{p}/0.3\,M_\mathrm{J})^2(a_\mathrm{NL}/a)^2$, such that we can neglect the first term, yielding
\beq
\gamma_\mathrm{NL} \simeq -\frac{1}{\tau_2}\exp\bigg(-\frac{\tau_2^2\Delta\omega^2}{(\pi/2)^2+1}\bigg(\frac{0.3 \, M_{\rm J}}{M_\mathrm{p}}\bigg)^2\bigg)\,.
\eeq

We note that the threshold amplitude $a_{\rm NL}$ inferred above is not necessarily the actual threshold amplitude for a nonlinear instability, because Figure 1 of \cite{essick:16} shows that even an off-resonance $0.1 \, M_{\rm J}$ planet excites a parent mode above the nonlinear threshold energy. However, the growth rate of the instability (and hence the amount of nonlinear damping) does apparently change rapidly with planet mass in this regime. A more accurate (but more complicated) model of nonlinear damping should incorporate the rapid increase in $|\gamma_{\rm NL}|$ at small mode amplitudes, the more gradual dependence $|\gamma_{\rm NL}| \propto |a|$ at intermediate amplitudes \citep{kumar:96,yu2020}, and the saturation $|\gamma_{\rm NL}| \sim 1/\tau_2$ at wave-breaking amplitudes. While such a model is beyond the scope of this work, it could significantly change both the width and depth of the resonant dips in Figure \ref{fig:RL_trapped}.


\bibliography{exobib,hotjupiter}{}

\begin{thebibliography}{}
\expandafter\ifx\csname natexlab\endcsname\relax\def\natexlab#1{#1}\fi
\providecommand{\url}[1]{\href{#1}{#1}}
\providecommand{\dodoi}[1]{doi:~\href{http://doi.org/#1}{\nolinkurl{#1}}}
\providecommand{\doeprint}[1]{\href{http://ascl.net/#1}{\nolinkurl{http://ascl.net/#1}}}
\providecommand{\doarXiv}[1]{\href{https://arxiv.org/abs/#1}{\nolinkurl{https://arxiv.org/abs/#1}}}

\bibitem[{{Auclair Desrotour} {et~al.}(2015){Auclair Desrotour}, {Mathis}, \&
  {Le Poncin-Lafitte}}]{auclair:15}
{Auclair Desrotour}, P., {Mathis}, S., \& {Le Poncin-Lafitte}, C. 2015, \aap,
  581, A118, \dodoi{10.1051/0004-6361/201526246}

\bibitem[{{Bailey} \& {Goodman}(2019)}]{bailey:19}
{Bailey}, A., \& {Goodman}, J. 2019, \mnras, 482, 1872,
  \dodoi{10.1093/mnras/sty2805}

\bibitem[{{Bakos} {et~al.}(2011){Bakos}, {Hartman}, {Torres}, {Latham},
  {Kov{\'a}cs}, {Noyes}, {Fischer}, {Johnson}, {Marcy}, {Howard}, {Kipping},
  {Esquerdo}, {Shporer}, {B{\'e}ky}, {Buchhave}, {Perumpilly}, {Everett},
  {Sasselov}, {Stefanik}, {L{\'a}z{\'a}r}, {Papp}, \& {S{\'a}ri}}]{Bakos2011}
{Bakos}, G.~{\'A}., {Hartman}, J., {Torres}, G., {et~al.} 2011, \apj, 742, 116,
  \dodoi{10.1088/0004-637X/742/2/116}

\bibitem[{{Barker}(2020)}]{barker2020}
{Barker}, A.~J. 2020, \mnras, 498, 2270, \dodoi{10.1093/mnras/staa2405}

\bibitem[{{Barker} \& {Ogilvie}(2010)}]{barker:10}
{Barker}, A.~J., \& {Ogilvie}, G.~I. 2010, \mnras, 404, 1849,
  \dodoi{10.1111/j.1365-2966.2010.16400.x}

\bibitem[{{Barker} \& {Ogilvie}(2011)}]{barker:11}
---. 2011, \mnras, 417, 745, \dodoi{10.1111/j.1365-2966.2011.19322.x}

\bibitem[{{Barros} {et~al.}(2016){Barros}, {Brown}, {H{\'e}brard}, {G{\'o}mez
  Maqueo Chew}, {Anderson}, {Boumis}, {Delrez}, {Hay}, {Lam}, {Llama}, {Lendl},
  {McCormac}, {Skiff}, {Smalley}, {Turner}, {Vanhuysse}, {Armstrong}, {Boisse},
  {Bouchy}, {Collier Cameron}, {Faedi}, {Gillon}, {Hellier}, {Jehin}, {Liakos},
  {Meaburn}, {Osborn}, {Pepe}, {Plauchu-Frayn}, {Pollacco}, {Queloz}, {Rey},
  {Spake}, {S{\'e}gransan}, {Triaud}, {Udry}, {Walker}, {Watson}, {West}, \&
  {Wheatley}}]{Barros2016}
{Barros}, S.~C.~C., {Brown}, D.~J.~A., {H{\'e}brard}, G., {et~al.} 2016, \aap,
  593, A113, \dodoi{10.1051/0004-6361/201526517}

\bibitem[{{Bouma} {et~al.}(2020){Bouma}, {Winn}, {Howard}, {Howell},
  {Isaacson}, {Knutson}, \& {Matson}}]{bouma2020}
{Bouma}, L.~G., {Winn}, J.~N., {Howard}, A.~W., {et~al.} 2020, \apjl, 893, L29,
  \dodoi{10.3847/2041-8213/ab8563}

\bibitem[{{Bouma} {et~al.}(2019){Bouma}, {Winn}, {Baxter}, {Bhatti}, {Dai},
  {Daylan}, {D{\'e}sert}, {Hill}, {Kane}, {Stassun}, {Villasenor}, {Ricker},
  {Vanderspek}, {Latham}, {Seager}, {Jenkins}, {Berta-Thompson}, {Col{\'o}n},
  {Fausnaugh}, {Glidden}, {Guerrero}, {Rodriguez}, {Twicken}, \&
  {Wohler}}]{bouma2019}
{Bouma}, L.~G., {Winn}, J.~N., {Baxter}, C., {et~al.} 2019, \aj, 157, 217,
  \dodoi{10.3847/1538-3881/ab189f}

\bibitem[{{Burkart} {et~al.}(2014){Burkart}, {Quataert}, \&
  {Arras}}]{burkart:14}
{Burkart}, J., {Quataert}, E., \& {Arras}, P. 2014, \mnras, 443, 2957,
  \dodoi{10.1093/mnras/stu1366}

\bibitem[{{Burkart} {et~al.}(2013){Burkart}, {Quataert}, {Arras}, \&
  {Weinberg}}]{burkart:13}
{Burkart}, J., {Quataert}, E., {Arras}, P., \& {Weinberg}, N.~N. 2013, \mnras,
  433, 332, \dodoi{10.1093/mnras/stt726}

\bibitem[{{Cheng} {et~al.}(2020){Cheng}, {Fuller}, {Guo}, {Lehman}, \&
  {Hambleton}}]{Cheng2020}
{Cheng}, S.~J., {Fuller}, J., {Guo}, Z., {Lehman}, H., \& {Hambleton}, K. 2020,
  \apj, 903, 122, \dodoi{10.3847/1538-4357/abb46d}

\bibitem[{{Collins} {et~al.}(2017){Collins}, {Kielkopf}, \&
  {Stassun}}]{collins2017}
{Collins}, K.~A., {Kielkopf}, J.~F., \& {Stassun}, K.~G. 2017, \aj, 153, 78,
  \dodoi{10.3847/1538-3881/153/2/78}

\bibitem[{{Delrez} {et~al.}(2018){Delrez}, {Madhusudhan}, {Lendl}, {Gillon},
  {Anderson}, {Neveu-VanMalle}, {Bouchy}, {Burdanov}, {Collier-Cameron},
  {Demory}, {Hellier}, {Jehin}, {Magain}, {Maxted}, {Queloz}, {Smalley}, \&
  {Triaud}}]{delrez2018}
{Delrez}, L., {Madhusudhan}, N., {Lendl}, M., {et~al.} 2018, \mnras, 474, 2334,
  \dodoi{10.1093/mnras/stx2896}

\bibitem[{{Duguid} {et~al.}(2020){Duguid}, {Barker}, \& {Jones}}]{duguid:20}
{Duguid}, C.~D., {Barker}, A.~J., \& {Jones}, C.~A. 2020, \mnras, 497, 3400,
  \dodoi{10.1093/mnras/staa2216}

\bibitem[{{Essick} \& {Weinberg}(2016)}]{essick:16}
{Essick}, R., \& {Weinberg}, N.~N. 2016, \apj, 816, 18,
  \dodoi{10.3847/0004-637X/816/1/18}

\bibitem[{{Fuller}(2017)}]{fuller2017}
{Fuller}, J. 2017, \mnras, 472, 1538, \dodoi{10.1093/mnras/stx2135}

\bibitem[{{Fuller} {et~al.}(2017){Fuller}, {Hambleton}, {Shporer}, {Isaacson},
  \& {Thompson}}]{fullerkic81:17}
{Fuller}, J., {Hambleton}, K., {Shporer}, A., {Isaacson}, H., \& {Thompson}, S.
  2017, \mnras, 472, L25, \dodoi{10.1093/mnrasl/slx130}

\bibitem[{{Fuller} \& {Lai}(2012)}]{fullerkoi54:12}
{Fuller}, J., \& {Lai}, D. 2012, \mnras, 420, 3126,
  \dodoi{10.1111/j.1365-2966.2011.20237.x}

\bibitem[{{Fuller} {et~al.}(2016){Fuller}, {Luan}, \& {Quataert}}]{fuller2016}
{Fuller}, J., {Luan}, J., \& {Quataert}, E. 2016, \mnras, 458, 3867,
  \dodoi{10.1093/mnras/stw609}

\bibitem[{{Gallet} {et~al.}(2017){Gallet}, {Bolmont}, {Mathis}, {Charbonnel},
  \& {Amard}}]{gallet:17}
{Gallet}, F., {Bolmont}, E., {Mathis}, S., {Charbonnel}, C., \& {Amard}, L.
  2017, \aap, 604, A112, \dodoi{10.1051/0004-6361/201730661}

\bibitem[{{Gillon} {et~al.}(2012){Gillon}, {Triaud}, {Fortney}, {Demory},
  {Jehin}, {Lendl}, {Magain}, {Kabath}, {Queloz}, {Alonso}, {Anderson},
  {Collier Cameron}, {Fumel}, {Hebb}, {Hellier}, {Lanotte}, {Maxted},
  {Mowlavi}, \& {Smalley}}]{gillon2012}
{Gillon}, M., {Triaud}, A.~H.~M.~J., {Fortney}, J.~J., {et~al.} 2012, \aap,
  542, A4, \dodoi{10.1051/0004-6361/201218817}

\bibitem[{{Gillon} {et~al.}(2013){Gillon}, {Anderson}, {Collier-Cameron},
  {Doyle}, {Fumel}, {Hellier}, {Jehin}, {Lendl}, {Maxted}, {Montalb{\'a}n},
  {Pepe}, {Pollacco}, {Queloz}, {S{\'e}gransan}, {Smith}, {Smalley},
  {Southworth}, {Triaud}, {Udry}, \& {West}}]{gillon2013}
{Gillon}, M., {Anderson}, D.~R., {Collier-Cameron}, A., {et~al.} 2013, \aap,
  552, A82, \dodoi{10.1051/0004-6361/201220561}

\bibitem[{{Gillon} {et~al.}(2014){Gillon}, {Anderson}, {Collier-Cameron},
  {Delrez}, {Hellier}, {Jehin}, {Lendl}, {Maxted}, {Pepe}, {Pollacco},
  {Queloz}, {S{\'e}gransan}, {Smith}, {Smalley}, {Southworth}, {Triaud},
  {Udry}, {Van Grootel}, \& {West}}]{gillon2014}
---. 2014, \aap, 562, L3, \dodoi{10.1051/0004-6361/201323014}

\bibitem[{{Goldreich} \& {Nicholson}(1989)}]{goldreich:89}
{Goldreich}, P., \& {Nicholson}, P.~D. 1989, \apj, 342, 1079,
  \dodoi{10.1086/167665}

\bibitem[{{Goldreich} \& {Soter}(1966)}]{goldreich1966}
{Goldreich}, P., \& {Soter}, S. 1966, \icarus, 5, 375,
  \dodoi{10.1016/0019-1035(66)90051-0}

\bibitem[{{Goldstein} \& {Townsend}(2020)}]{goldstein2020}
{Goldstein}, J., \& {Townsend}, R.~H.~D. 2020, \apj, 899, 116,
  \dodoi{10.3847/1538-4357/aba748}

\bibitem[{{Goodman} \& {Dickson}(1998)}]{goodman:98}
{Goodman}, J., \& {Dickson}, E.~S. 1998, \apj, 507, 938, \dodoi{10.1086/306348}

\bibitem[{{Guenel} {et~al.}(2016){Guenel}, {Baruteau}, {Mathis}, \&
  {Rieutord}}]{guenel:16}
{Guenel}, M., {Baruteau}, C., {Mathis}, S., \& {Rieutord}, M. 2016, \aap, 589,
  A22, \dodoi{10.1051/0004-6361/201527621}

\bibitem[{{Hambleton} {et~al.}(2018){Hambleton}, {Fuller}, {Thompson},
  {Pr{\v{s}}a}, {Kurtz}, {Shporer}, {Isaacson}, {Howard}, {Endl}, {Cochran}, \&
  {Murphy}}]{hambleton:18}
{Hambleton}, K., {Fuller}, J., {Thompson}, S., {et~al.} 2018, \mnras, 473,
  5165, \dodoi{10.1093/mnras/stx2673}

\bibitem[{{Hamer} \& {Schlaufman}(2019)}]{hamer:20a}
{Hamer}, J.~H., \& {Schlaufman}, K.~C. 2019, \aj, 158, 190,
  \dodoi{10.3847/1538-3881/ab3c56}

\bibitem[{{Hamer} \& {Schlaufman}(2020)}]{hamer:20b}
---. 2020, \aj, 160, 138, \dodoi{10.3847/1538-3881/aba74f}

\bibitem[{{Hebb} {et~al.}(2009){Hebb}, {Collier-Cameron}, {Loeillet},
  {Pollacco}, {H{\'e}brard}, {Street}, {Bouchy}, {Stempels}, {Moutou},
  {Simpson}, {Udry}, {Joshi}, {West}, {Skillen}, {Wilson}, {McDonald},
  {Gibson}, {Aigrain}, {Anderson}, {Benn}, {Christian}, {Enoch}, {Haswell},
  {Hellier}, {Horne}, {Irwin}, {Lister}, {Maxted}, {Mayor}, {Norton}, {Parley},
  {Pont}, {Queloz}, {Smalley}, \& {Wheatley}}]{Hebb2009}
{Hebb}, L., {Collier-Cameron}, A., {Loeillet}, B., {et~al.} 2009, \apj, 693,
  1920, \dodoi{10.1088/0004-637X/693/2/1920}

\bibitem[{{Hebb} {et~al.}(2010){Hebb}, {Collier-Cameron}, {Triaud}, {Lister},
  {Smalley}, {Maxted}, {Hellier}, {Anderson}, {Pollacco}, {Gillon}, {Queloz},
  {West}, {Bentley}, {Enoch}, {Haswell}, {Horne}, {Mayor}, {Pepe}, {Segransan},
  {Skillen}, {Udry}, \& {Wheatley}}]{Hebb2010}
{Hebb}, L., {Collier-Cameron}, A., {Triaud}, A.~H.~M.~J., {et~al.} 2010, \apj,
  708, 224, \dodoi{10.1088/0004-637X/708/1/224}

\bibitem[{{Hellier} {et~al.}(2009){Hellier}, {Anderson}, {Collier Cameron},
  {Gillon}, {Hebb}, {Maxted}, {Queloz}, {Smalley}, {Triaud}, {West}, {Wilson},
  {Bentley}, {Enoch}, {Horne}, {Irwin}, {Lister}, {Mayor}, {Parley}, {Pepe},
  {Pollacco}, {Segransan}, {Udry}, \& {Wheatley}}]{hellier2009}
{Hellier}, C., {Anderson}, D.~R., {Collier Cameron}, A., {et~al.} 2009, \nat,
  460, 1098, \dodoi{10.1038/nature08245}

\bibitem[{{Hellier} {et~al.}(2011){Hellier}, {Anderson}, {Collier Cameron},
  {Gillon}, {Jehin}, {Lendl}, {Maxted}, {Pepe}, {Pollacco}, {Queloz},
  {S{\'e}gransan}, {Smalley}, {Smith}, {Southworth}, {Triaud}, {Udry}, \&
  {West}}]{Hellier2011}
---. 2011, \aap, 535, L7, \dodoi{10.1051/0004-6361/201117081}

\bibitem[{{Hod{\v{z}}i{\'c}} {et~al.}(2018){Hod{\v{z}}i{\'c}}, {Triaud},
  {Anderson}, {Bouchy}, {Collier Cameron}, {Delrez}, {Gillon}, {Hellier},
  {Jehin}, {Lendl}, {Maxted}, {Pepe}, {Pollacco}, {Queloz}, {S{\'e}gransan},
  {Smalley}, {Udry}, \& {West}}]{Hodzic2018}
{Hod{\v{z}}i{\'c}}, V., {Triaud}, A. H.~M.~J., {Anderson}, D.~R., {et~al.}
  2018, \mnras, 481, 5091, \dodoi{10.1093/mnras/sty2512}

\bibitem[{{Ivanov} {et~al.}(2013){Ivanov}, {Papaloizou}, \&
  {Chernov}}]{ivanov:13}
{Ivanov}, P.~B., {Papaloizou}, J.~C.~B., \& {Chernov}, S.~V. 2013, \mnras, 432,
  2339, \dodoi{10.1093/mnras/stt595}

\bibitem[{{Krishnamurthi} {et~al.}(1997){Krishnamurthi}, {Pinsonneault},
  {Barnes}, \& {Sofia}}]{Krishnamurthi1997}
{Krishnamurthi}, A., {Pinsonneault}, M.~H., {Barnes}, S., \& {Sofia}, S. 1997,
  \apj, 480, 303, \dodoi{10.1086/303958}

\bibitem[{{Kumar} \& {Goodman}(1996)}]{kumar:96}
{Kumar}, P., \& {Goodman}, J. 1996, \apj, 466, 946, \dodoi{10.1086/177565}

\bibitem[{{Lainey} {et~al.}(2020){Lainey}, {Gomez Casajus}, {Fuller},
  {Zannoni}, \& {Tortora}}]{Lainey2020}
{Lainey}, V., {Gomez Casajus}, L., {Fuller}, J., {Zannoni}, M., \& {Tortora},
  P. e.~a. 2020, Nature Astronomy

\bibitem[{{Lainey} {et~al.}(2017){Lainey}, {Jacobson}, {Tajeddine}, {Cooper},
  {Murray}, {Robert}, {Tobie}, {Guillot}, {Mathis}, {Remus}, {Desmars},
  {Arlot}, {De Cuyper}, {Dehant}, {Pascu}, {Thuillot}, {Le Poncin-Lafitte}, \&
  {Zahn}}]{lainey:17}
{Lainey}, V., {Jacobson}, R.~A., {Tajeddine}, R., {et~al.} 2017, Icarus, 281,
  286, \dodoi{10.1016/j.icarus.2016.07.014}

\bibitem[{{Lee} \& {Chiang}(2017)}]{lee:17}
{Lee}, E.~J., \& {Chiang}, E. 2017, \apj, 842, 40,
  \dodoi{10.3847/1538-4357/aa6fb3}

\bibitem[{{Mancini} {et~al.}(2013){Mancini}, {Ciceri}, {Chen}, {Tregloan-Reed},
  {Fortney}, {Southworth}, {Tan}, {Burgdorf}, {Calchi Novati}, {Dominik},
  {Fang}, {Finet}, {Gerner}, {Hardis}, {Hinse}, {J{\o}rgensen}, {Liebig},
  {Nikolov}, {Ricci}, {Sch{\"a}fer}, {Sch{\"o}nebeck}, {Skottfelt}, {Wertz},
  {Alsubai}, {Bozza}, {Browne}, {Dodds}, {Gu}, {Harps{\o}e}, {Henning},
  {Hundertmark}, {Jessen-Hansen}, {Kains}, {Kerins}, {Kjeldsen}, {Lund},
  {Lundkvist}, {Madhusudhan}, {Mathiasen}, {Penny}, {Prof}, {Rahvar}, {Sahu},
  {Scarpetta}, {Snodgrass}, \& {Surdej}}]{mancini2013}
{Mancini}, L., {Ciceri}, S., {Chen}, G., {et~al.} 2013, \mnras, 436, 2,
  \dodoi{10.1093/mnras/stt1394}

\bibitem[{{Mannaday} {et~al.}(2020){Mannaday}, {Thakur}, {Jiang}, {Sahu},
  {Joshi}, {Pandey}, {Joshi}, {Yadav}, {Su}, {Sariya}, {Yeh}, {Griv},
  {Mkrtichian}, {Shlyapnikov}, {Moskvin}, {Ignatov}, {Va{\v{n}}ko}, \&
  {P{\"u}sk{\"u}ll{\"u}}}]{mannaday2020}
{Mannaday}, V.~K., {Thakur}, P., {Jiang}, I.-G., {et~al.} 2020, \aj, 160, 47,
  \dodoi{10.3847/1538-3881/ab9818}

\bibitem[{{Mathis}(2015)}]{mathis:15}
{Mathis}, S. 2015, \aap, 580, L3, \dodoi{10.1051/0004-6361/201526472}

\bibitem[{{Mathis} {et~al.}(2016){Mathis}, {Auclair-Desrotour}, {Guenel},
  {Gallet}, \& {Le Poncin-Lafitte}}]{mathis:16}
{Mathis}, S., {Auclair-Desrotour}, P., {Guenel}, M., {Gallet}, F., \& {Le
  Poncin-Lafitte}, C. 2016, \aap, 592, A33, \dodoi{10.1051/0004-6361/201527545}

\bibitem[{{Millholland} \& {Spalding}(2020)}]{millholland:20}
{Millholland}, S.~C., \& {Spalding}, C. 2020, \apj, 905, 71,
  \dodoi{10.3847/1538-4357/abc4e5}

\bibitem[{{Oberst} {et~al.}(2017){Oberst}, {Rodriguez}, {Col{\'o}n},
  {Angerhausen}, {Bieryla}, {Ngo}, {Stevens}, {Stassun}, {Gaudi}, {Pepper},
  {Penev}, {Mawet}, {Latham}, {Heintz}, {Osei}, {Collins}, {Kielkopf},
  {Visgaitis}, {Reed}, {Escamilla}, {Yazdi}, {McLeod}, {Lunsford}, {Spencer},
  {Joner}, {Gregorio}, {Gaillard}, {Matt}, {Dumont}, {Stephens}, {Cohen},
  {Jensen}, {Calchi Novati}, {Bozza}, {Labadie-Bartz}, {Siverd}, {Lund},
  {Beatty}, {Eastman}, {Penny}, {Manner}, {Zambelli}, {Fulton}, {Stockdale},
  {DePoy}, {Marshall}, {Pogge}, {Gould}, {Trueblood}, \&
  {Trueblood}}]{oberst2017}
{Oberst}, T.~E., {Rodriguez}, J.~E., {Col{\'o}n}, K.~D., {et~al.} 2017, \aj,
  153, 97, \dodoi{10.3847/1538-3881/153/3/97}

\bibitem[{{O'Donovan} {et~al.}(2007){O'Donovan}, {Charbonneau}, {Bakos},
  {Mandushev}, {Dunham}, {Brown}, {Latham}, {Torres}, {Sozzetti}, {Kov{\'a}cs},
  {Everett}, {Baliber}, {Hidas}, {Esquerdo}, {Rabus}, {Deeg}, {Belmonte},
  {Hillenbrand}, \& {Stefanik}}]{odonovan2007}
{O'Donovan}, F.~T., {Charbonneau}, D., {Bakos}, G.~{\'A}., {et~al.} 2007,
  \apjl, 663, L37, \dodoi{10.1086/519793}

\bibitem[{{Ogilvie}(2013)}]{ogilvie:13}
{Ogilvie}, G.~I. 2013, \mnras, 429, 613, \dodoi{10.1093/mnras/sts362}

\bibitem[{{Owen} \& {Lai}(2018)}]{owen:18}
{Owen}, J.~E., \& {Lai}, D. 2018, \mnras, 479, 5012,
  \dodoi{10.1093/mnras/sty1760}

\bibitem[{{Papaloizou} \& {Ivanov}(2010)}]{papaloizou:10}
{Papaloizou}, J.~C.~B., \& {Ivanov}, P.~B. 2010, \mnras, 407, 1631,
  \dodoi{10.1111/j.1365-2966.2010.17011.x}

\bibitem[{{Patra} {et~al.}(2017){Patra}, {Winn}, {Holman}, {Yu}, {Deming}, \&
  {Dai}}]{patra2017}
{Patra}, K.~C., {Winn}, J.~N., {Holman}, M.~J., {et~al.} 2017, \aj, 154, 4,
  \dodoi{10.3847/1538-3881/aa6d75}

\bibitem[{{Patra} {et~al.}(2020){Patra}, {Winn}, {Holman}, {Gillon},
  {Burdanov}, {Jehin}, {Delrez}, {Pozuelos}, {Barkaoui}, {Benkhaldoun},
  {Narita}, {Fukui}, {Kusakabe}, {Kawauchi}, {Terada}, {Bouma}, {Weinberg}, \&
  {Broome}}]{patra2020}
---. 2020, \aj, 159, 150, \dodoi{10.3847/1538-3881/ab7374}

\bibitem[{{Paxton} {et~al.}(2011){Paxton}, {Bildsten}, {Dotter}, {Herwig},
  {Lesaffre}, \& {Timmes}}]{paxton2011}
{Paxton}, B., {Bildsten}, L., {Dotter}, A., {et~al.} 2011, \apjs, 192, 3,
  \dodoi{10.1088/0067-0049/192/1/3}

\bibitem[{{Paxton} {et~al.}(2013){Paxton}, {Cantiello}, {Arras}, {Bildsten},
  {Brown}, {Dotter}, {Mankovich}, {Montgomery}, {Stello}, {Timmes}, \&
  {Townsend}}]{paxton2013}
{Paxton}, B., {Cantiello}, M., {Arras}, P., {et~al.} 2013, \apjs, 208, 4,
  \dodoi{10.1088/0067-0049/208/1/4}

\bibitem[{{Paxton} {et~al.}(2015){Paxton}, {Marchant}, {Schwab}, {Bauer},
  {Bildsten}, {Cantiello}, {Dessart}, {Farmer}, {Hu}, {Langer}, {Townsend},
  {Townsley}, \& {Timmes}}]{paxton2015}
{Paxton}, B., {Marchant}, P., {Schwab}, J., {et~al.} 2015, \apjs, 220, 15,
  \dodoi{10.1088/0067-0049/220/1/15}

\bibitem[{{Paxton} {et~al.}(2018){Paxton}, {Schwab}, {Bauer}, {Bildsten},
  {Blinnikov}, {Duffell}, {Farmer}, {Goldberg}, {Marchant}, {Sorokina},
  {Thoul}, {Townsend}, \& {Timmes}}]{paxton2018}
{Paxton}, B., {Schwab}, J., {Bauer}, E.~B., {et~al.} 2018, \apjs, 234, 34,
  \dodoi{10.3847/1538-4365/aaa5a8}

\bibitem[{{Paxton} {et~al.}(2019){Paxton}, {Smolec}, {Schwab}, {Gautschy},
  {Bildsten}, {Cantiello}, {Dotter}, {Farmer}, {Goldberg}, {Jermyn}, {Kanbur},
  {Marchant}, {Thoul}, {Townsend}, {Wolf}, {Zhang}, \& {Timmes}}]{paxton2019}
{Paxton}, B., {Smolec}, R., {Schwab}, J., {et~al.} 2019, \apjs, 243, 10,
  \dodoi{10.3847/1538-4365/ab2241}

\bibitem[{{Penev} {et~al.}(2018){Penev}, {Bouma}, {Winn}, \&
  {Hartman}}]{penev2018}
{Penev}, K., {Bouma}, L.~G., {Winn}, J.~N., \& {Hartman}, J.~D. 2018, \aj, 155,
  165, \dodoi{10.3847/1538-3881/aaaf71}

\bibitem[{{Penev} {et~al.}(2014){Penev}, {Zhang}, \& {Jackson}}]{penev:14}
{Penev}, K., {Zhang}, M., \& {Jackson}, B. 2014, \pasp, 126, 553,
  \dodoi{10.1086/677042}

\bibitem[{{Penev} {et~al.}(2016){Penev}, {Hartman}, {Bakos}, {Ciceri}, {Brahm},
  {Bayliss}, {Bento}, {Jord{\'a}n}, {Csubry}, {Bhatti}, {de Val-Borro},
  {Espinoza}, {Zhou}, {Mancini}, {Rabus}, {Suc}, {Henning}, {Schmidt}, {Noyes},
  {L{\'a}z{\'a}r}, {Papp}, \& {S{\'a}ri}}]{penev2016}
{Penev}, K., {Hartman}, J.~D., {Bakos}, G.~{\'A}., {et~al.} 2016, \aj, 152,
  127, \dodoi{10.3847/0004-6256/152/5/127}

\bibitem[{{Sasselov}(2003)}]{sasselov2003}
{Sasselov}, D.~D. 2003, \apj, 596, 1327, \dodoi{10.1086/378145}

\bibitem[{{Skumanich}(1972)}]{skumanich1972}
{Skumanich}, A. 1972, \apj, 171, 565, \dodoi{10.1086/151310}

\bibitem[{{Southworth} {et~al.}(2019){Southworth}, {Dominik}, {J{\o}rgensen},
  {Andersen}, {Bozza}, {Burgdorf}, {D'Ago}, {Dib}, {Figuera Jaimes}, {Fujii},
  {Gill}, {Haikala}, {Hinse}, {Hundertmark}, {Khalouei}, {Korhonen},
  {Longa-Pe{\~n}a}, {Mancini}, {Peixinho}, {Rabus}, {Rahvar}, {Sajadian},
  {Skottfelt}, {Snodgrass}, {Spyratos}, {Tregloan-Reed}, {Unda-Sanzana}, \&
  {von Essen}}]{Southworth2019}
{Southworth}, J., {Dominik}, M., {J{\o}rgensen}, U.~G., {et~al.} 2019, \mnras,
  490, 4230, \dodoi{10.1093/mnras/stz2602}

\bibitem[{{Stassun} {et~al.}(2017){Stassun}, {Collins}, \&
  {Gaudi}}]{stassun2017}
{Stassun}, K.~G., {Collins}, K.~A., \& {Gaudi}, B.~S. 2017, \aj, 153, 136,
  \dodoi{10.3847/1538-3881/aa5df3}

\bibitem[{{Torres} {et~al.}(2008){Torres}, {Winn}, \& {Holman}}]{torres2008}
{Torres}, G., {Winn}, J.~N., \& {Holman}, M.~J. 2008, \apj, 677, 1324,
  \dodoi{10.1086/529429}

\bibitem[{{Townsend} {et~al.}(2018){Townsend}, {Goldstein}, \&
  {Zweibel}}]{townsend2018}
{Townsend}, R.~H.~D., {Goldstein}, J., \& {Zweibel}, E.~G. 2018, \mnras, 475,
  879, \dodoi{10.1093/mnras/stx3142}

\bibitem[{{Townsend} \& {Teitler}(2013)}]{townsend2013}
{Townsend}, R.~H.~D., \& {Teitler}, S.~A. 2013, \mnras, 435, 3406,
  \dodoi{10.1093/mnras/stt1533}

\bibitem[{{Turner} {et~al.}(2016){Turner}, {Anderson}, {Collier Cameron},
  {Delrez}, {Evans}, {Gillon}, {Hellier}, {Jehin}, {Lendl}, {Maxted}, {Pepe},
  {Pollacco}, {Queloz}, {S{\'e}gransan}, {Smalley}, {Smith}, {Triaud}, {Udry},
  \& {West}}]{turner2016}
{Turner}, O.~D., {Anderson}, D.~R., {Collier Cameron}, A., {et~al.} 2016,
  \pasp, 128, 064401, \dodoi{10.1088/1538-3873/128/964/064401}

\bibitem[{{Vidal} \& {Barker}(2020)}]{vidal:20}
{Vidal}, J., \& {Barker}, A.~J. 2020, \apjl, 888, L31,
  \dodoi{10.3847/2041-8213/ab6219}

\bibitem[{{Weinberg} {et~al.}(2012){Weinberg}, {Arras}, {Quataert}, \&
  {Burkart}}]{weinberg:12}
{Weinberg}, N.~N., {Arras}, P., {Quataert}, E., \& {Burkart}, J. 2012, \apj,
  751, 136, \dodoi{10.1088/0004-637X/751/2/136}

\bibitem[{{Weinberg} {et~al.}(2017){Weinberg}, {Sun}, {Arras}, \&
  {Essick}}]{weinberg2017}
{Weinberg}, N.~N., {Sun}, M., {Arras}, P., \& {Essick}, R. 2017, \apjl, 849,
  L11, \dodoi{10.3847/2041-8213/aa9113}

\bibitem[{{Witte} \& {Savonije}(1999)}]{witte:99}
{Witte}, M.~G., \& {Savonije}, G.~J. 1999, \aap, 350, 129

\bibitem[{{Witte} \& {Savonije}(2001)}]{witte:01}
---. 2001, \aap, 366, 840, \dodoi{10.1051/0004-6361:20000245}

\bibitem[{{Yu} {et~al.}(2020){Yu}, {Weinberg}, \& {Fuller}}]{yu2020}
{Yu}, H., {Weinberg}, N.~N., \& {Fuller}, J. 2020, \mnras, 496, 5482,
  \dodoi{10.1093/mnras/staa1858}

\bibitem[{{Zahn}(1975)}]{zahn:75}
{Zahn}, J.-P. 1975, \aap, 41, 329

\bibitem[{{Zanazzi} \& {Wu}(2021)}]{Zanazzi2021}
{Zanazzi}, J.~J., \& {Wu}, Y. 2021, \aj, 161, 263,
  \dodoi{10.3847/1538-3881/abf097}

\bibitem[{{Zhu} \& {Dong}(2021)}]{zhu2021}
{Zhu}, W., \& {Dong}, S. 2021, arXiv e-prints, arXiv:2103.02127.
\newblock \doarXiv{2103.02127}

\end{thebibliography}
\bibliographystyle{aasjournal}



\end{document}